\documentclass[epjST]{svjour}
\usepackage{graphics}
\usepackage{xcolor}
\usepackage{slashed}
\usepackage{mathrsfs}
\usepackage{amsmath}
\usepackage{hyperref}
\usepackage{cite}

\begin{document}
\title{Structure-dependent QED effects in exclusive $B$-meson decays\thanks{MITP-23-072, SI-HEP-2023-35, P3H-23-099} }

\author{%
	Philipp B\"oer\inst{1}\fnmsep\thanks{\email{pboeer@uni-mainz.de}} 
	\and Thorsten Feldmann \inst{2}\fnmsep\thanks{\email{thorsten.feldmann@uni-siegen.de}} }
\institute{%
PRISMA$^+$ Cluster of Excellence \& Mainz Institute for Theoretical Physics, \\ Johannes Gutenberg Universit\"at, Staudingerweg 9, 55128 Mainz, Germany
	\and
	Theoretische Physik 1, Center for Particle Physics Siegen, \\ Universit\"at Siegen, Walter-Flex-Stra\ss{}e 3, 57068  Siegen,  Germany
 }

\abstract{%
    We review recent progress in the computation of structure-dependent QED corrections to exclusive $B$ decays in the factorization approach.
}
\maketitle

\tableofcontents

\section{Introduction}\label{sec1}

Weak decays of $b$-quarks continue to play a very important role for the exploration of the flavour sector in the Standard Model (SM) and for the search of new particles and interactions from physics beyond the SM (BSM).
Our present knowledge is based on 
a huge amount of experimental data from $e^+e^-$  ``B-factories'' and hadron colliders, 
supplemented with refined theoretical methods and computational tools to control radiative QCD corrections and hadronic input parameters
(for reviews, see e.g.~\cite{LHCb:2012myk,Belle-II:2018jsg} and references therein).
Nowadays, measurements of flavour processes have reached an astonishing level of precision, and -- in most cases -- experimental results 
turn out to be perfectly in line with theoretical expectations within the SM. On the other hand, a handful of deviations between theory and experiment 
have been reported for certain flavour observables in recent years, which may point to BSM contributions in these decays (a recent review on these so-called
``flavour anomalies'' can be found in~\cite{Albrecht:2021tul}; see also~\cite{Capdevila:2023yhq} in this volume). 
In this situation, further improvements in precision on the experimental and theoretical side are indispensable. 
In particular, at some level, QED effects become relevant and have to be included in a systematic manner. 
In particular, compared to QCD corrections, the inclusion of QED radiation can lead to qualitatively new effects which in some cases can mimic short-distance 
new physics. Important examples are logarithmic dependencies on charged-lepton masses which violate flavour universality, or new sources of isospin violation from the different quark charges.
While this clearly affects the sensitivity to BSM physics, it also concerns practical issues 
related to the consistent implementation of photon radiation into
the phenomenological analyses of experimental data. Last but not least, the interplay between electromagnetic and strong interactions at low energies puts 
new theoretical challenges for a proper field-theoretical formulation. 

The purpose of this review is to summarize the current status of our theoretical understanding of electromagnetic corrections in exclusive $B$-meson decays into energetic particles.
Here, we will focus on those effects where the virtual photons resolve the hadronic structure in the initial and final state.
Such so-called structure-dependent corrections go beyond the traditional treatment of soft QED effects in $B$ decays, 
and are also not included in typical Monte Carlo implementations (such as {\sc PHOTOS}~\cite{Golonka:2005pn}) 
of electromagnetic radiation in experimental analyses. Both assume pointlike photon couplings to charged hadrons up to energy scales of $\mathcal{O}(m_B)$. 
However, as the spatial extension of hadrons is of order $1/\Lambda $ (with $\Lambda$ being a hadronic scale of order of a few hundred MeV),
photons with wavelengths as small as $1/m_B$ can resolve the partonic substructure, 
which requires a more sophisticated theoretical description.

The outline of this review article is as follows. In the next section we start with a overview of the general 
theoretical background, explaining the different energy scales of the problem and the resulting ingredients of the relevant effective field theories.
In Sec.~\ref{sec:QEDF} we focus on the specific issues that have to be accounted for in QED factorization and summarize the results that have been
obtained in the literature for leptonic, semi-leptonic and non-leptonic decay modes. As an important new feature of QED factorization theorems, we 
also discuss the properties of QED-generalized light-cone distribution amplitudes for light and heavy mesons.
Sec.~\ref{sec:rapidity} is devoted to the treatment of rapidity logarithms which have been analyzed in $B_s \to \mu^+\mu^-$ and $B^- \to \mu^- \bar \nu_\ell$ decays.
We close with a brief summary.

\section[Theoretical background: factorization and effective field theories]{Theoretical background: \newline \phantom{2} factorization and effective field theories}

The modern theoretical description of exclusive 
$b$-hadron decay amplitudes is based on the concept of \emph{factorization}, i.e.\ the separation of quantum corrections from different mass and energy scales. 
Concerning QED effects, this clear separation of energy scales is pivotal to separate very low-energy photons with pointlike couplings to mesons from photons with larger energy (smaller wavelength), which resolve their partonic substructure.
Mathematically, this separation is encoded in \emph{factorization theorems} for the corresponding decay amplitudes which in QCD first have been systematically studied for 
non-leptonic $B$-meson decays in \cite{Beneke:1999br,Beneke:2000ry,Beneke:2001ev}. 
In perturbation theory, an intuitive approach is to approximate the relevant Feynman integrals
by expanding in ratios of small and large masses or energies/momenta 
for specific \emph{momentum regions} of the loop momenta \cite{Beneke:1997zp}. 
Such expansions can be made explicit at the Lagrangian level, leading to the concept of \emph{effective field theories} (EFTs) which underlies the modern formulation of factorization theorems.
 Here, the quantum effects from large energy scales are encoded in  Wilson coefficients (or, more generally, coefficient \emph{functions}) which are identified with the short-distance functions in the factorization theorem. On the other hand, the dynamics at low energy scales is reproduced in terms of effective operators, whose matrix elements determine the long-distance parameters/functions. 
 One of the advantages of the EFT formulation lies in the fact that one can systematically keep track of fundamental symmetries of the underlying theory, as well as of (new) approximate symmetries of the low-energy expansion. Moreover, the \emph{renormalization-group equations} (RGEs) for the effective operators can be used to systematically sum large logarithms of hierarchical scale ratios to all orders in perturbation theory.
 For phenomenological applications in $b$-physics, one is then left with a number of hadronic matrix elements of EFT operators, which have to be quantified by means of non-perturbative 
 QCD methods. The state-of-the-art of such calculations is reviewed in \cite{Tsang:2023nay} for the lattice-QCD approach, and in \cite{Khodjamirian:2023wol} for QCD sum-rule methods, as part of this special volume.

Within the SM, transitions between different quark flavours are mediated by the gauge bosons of the weak interactions $W^\pm, Z^0$ with masses of order 100~GeV. 
As the energy and momentum transfer in weak $b$-quark decays are bounded by the $b$-quark mass, $m_b \ll M_{W,Z}$, one can describe the relevant interactions in
terms of a \emph{weak effective Hamiltonian}, which takes the generic form
\begin{eqnarray}
  H_{\rm ew}^{\rm eff}(x) &=& \sum_i C_i(\mu) \, {\cal O}_i(x) \,,
  \label{eq:intro:Heff}
\end{eqnarray}
where ${\cal O}_i(x)$ are \emph{local} operators constructed from products of field operators for light quarks ($b,c,s,d,u$), leptons, gluons and photons.
The set of these operators is constrained by the SM symmetries and can be truncated by including only operators up to a given canonical mass dimension.
Examples will be discussed in the specific applications below, for comprehensive reviews see \cite{Buchalla:1995vs,Chetyrkin:1996vx}.
The functions $C_i(\mu)$ are referred to as (electroweak) Wilson coefficients, and they contain 
the dynamical effects of short-distance quantum fluctuations above a given factorization scale $\mu$. The scale-dependence of the $C_i(\mu)$ is governed by 
the RGEs, 
\begin{eqnarray}
  \frac{d}{d\ln\mu} \, C_i(\mu) = \gamma_{ij}(\mu) \, C_j(\mu) \,,
\end{eqnarray}
where $\gamma_{ij}$ is the anomalous-dimension matrix. 
The entries $\gamma_{ij}$ can be computed perturbatively, incorporating radiative corrections at energy scales $\mu$ inbetween the electroweak and $b$-quark scale. 
Solving the above RGE results in very precise predictions for the Wilson coefficients $C_i(\mu \sim m_b)$ within the SM. Moreover, assuming modifications of the electroweak sector from
``new physics'' above the electroweak scale, the framework can easily be extended by enlarging the operator basis and allowing for modifications of the values of the Wilson coefficients.
In this way the Wilson coefficients contain \emph{all} the dependence on the high-energy parameters ($M_{W,Z}$ etc.) in the SM or its possible new-physics extensions.

After factoring out the short-distance Wilson coefficients, the calculation of transition amplitudes for a given flavour process requires to compute hadronic matrix-elements of the effective operators,
$$
 \langle {\cal O}_i \rangle \equiv \langle \mbox{final state}| {\cal O}_i |\mbox{initial state}\rangle \,, 
$$
which -- by construction -- are independent of the short-distance parameters. Instead, they now depend on the factorization scale, $\langle {\cal O}_i \rangle = \langle {\cal O}_i \rangle(\mu)$, which cancels the scale dependence of the Wilson coefficients in~\eqref{eq:intro:Heff}. 
For the specific case of exclusive $b$-hadron decays, one usually computes these matrix elements at the scale \mbox{$\mu= \mu_b \sim m_b \approx 5$~GeV}.
From the hadronic perspective this now provides the \emph{largest} scale in the dynamics entering the matrix elements $\langle {\cal O}_i \rangle$,
and is thus 
usually referred to as the \emph{hard scale}. On the other hand, the low scale $\Lambda \equiv \Lambda_{\rm had.} \sim$~a~few 100~MeV, 
is set by the long-distance fluctuations in hadronic bound states in QCD, which is usually referred to as the \emph{soft scale} in the factorization theorem. At such low energy scales QCD is strongly coupled and is no longer accessible in perturbation theory.

Factorization of the hard and soft scale is achieved through an expansion in the small scale ratio $\Lambda/m_b \ll 1$, and is essential for a clear separation of perturbatively calculable short distance objects from non-perturbative hadronic quantities. To construct this expansion, the first key observation is to realize that -- from the low-energy perspective -- the $b$-quark in a $b$-hadron bound state can be approximated by a static colour source, moving almost with the same 4-velocity $v^\mu$ as the $b$-hadron itself,
\begin{eqnarray}
	p_b^\mu = m_b \, v^\mu +  k^\mu \,, \qquad | k^\mu| \sim {\cal O}(\Lambda) \ll m_b \,.
\end{eqnarray}
This situation can be described by an effective Lagrangian,
which represents the leading term in the so-called \emph{heavy-quark effective theory} (HQET, see \cite{Neubert:1993mb} and references therein),
\begin{eqnarray}
	 {\cal L}_{\rm HQET}^{(0)}(x) &=& \bar h_v(x) \, (i v \cdot D) \, h_v(x) \,, \qquad \slashed v \, h_v(x) = h_v(x) \,.
\end{eqnarray}
Here, the field operators $h_v(x)$ describe the fluctuations related to the residual momentum $k^\mu$ around the static limit, projected onto the relevant (large) spinor components.

In exclusive $b$-decays, we distinguish two fundamentally different kinematic situations:
\begin{itemize}
	\item[(i)] The energy transfer of the decaying $b$-quark to the final-state quarks and gluons is small. This happens, for instance, in semi-leptonic decays $B \to D \ell \bar \nu_\ell$, and also in $B \to \pi \ell \bar \nu_\ell$, if the invariant mass of the lepton pair is large. In that case the EFT construction will be in terms of soft fields only, i.e.\ $h_v(x)$ for the heavy $b$- and $c$-quarks, and the usual light-quark and gluon fields.
    The effective Lagrangian in this case is given in terms of a series of \emph{local} operators of increasing canonical mass dimension. The hadronic matrix elements of these operators define simple hadronic quantities, like mesonic decay constants or (generalized) transition form factors in HQET.

\vspace{0.2em}
 
	\item[(ii)] In contrast, quarks and gluons that end up in the final state might have large energies, i.e.\ they are Lorentz-boosted with respect to the $b$-quark rest frame. This happens, for instance, for charmless non-leptonic decays ($B \to \pi \pi$ etc.) and for semi-leptonic $B \to \pi \ell \bar \nu_\ell$ decays at small invariant lepton mass. 
    In that case, the effective description must -- in addition to soft modes -- also incorporate energetic ``collinear'' degrees of freedom, and is referred to as \emph{soft-collinear effective theory} (SCET \cite{Bauer:2000yr,Bauer:2001yt,Beneke:2002ph,Bauer:2002nz,Beneke:2002ni}).
    For a pedagogical introduction and further references we refer to the comprehensive review in~\cite{Becher:2014oda}.
\end{itemize}
In the following, we will focus on the second case, i.e.\ transitions involving light and energetic final-state particles.

\subsection{Ingredients of the SCET Lagrangian and currents}
 
In general, the SCET Lagrangian involves light-quark and gluon fields for soft momentum modes, 
 as well as for \emph{collinear} modes in each relevant direction (one direction for $B\to \pi\ell\nu$, two directions for $B\to\pi\pi$ etc.),
    \begin{align}
     {\cal L}_{\rm SCET} & = 
     {\cal L}[h_v,q_s,A_s; q_{c_i},A_{c_i}] \,.
    \end{align}
    As a consequence of the large Lorentz boost, the interaction Lagrangian contains non-local operators that have to be multipole-expanded with respect to the typical wave-lengths of collinear and soft field modes in different directions \cite{Beneke:2002ni}.
    Moreover, in order to achieve manifest invariance with respect to independent  
    gauge transformations in the soft and collinear sectors, soft and collinear Wilson lines appear at soft-collinear vertices. Finally, 
    by a suitable field redefinition one can remove the interactions between soft and collinear degrees of freedom from the leading power Lagrangian. This decoupling transformation is implicitly understood in the following.

A further complication in exclusive $b$-quark decays stems from the fact that interactions of soft and collinear fields happen via the exchange of so-called \emph{hard-collinear} field quanta 
    with intermediate virtualities, $(k_s + p_c)^2 \sim {\cal O}(|k_s| \, E_c) \sim 
    {\cal O}(\Lambda \, m_b)$, where $E_c$ is the large energy of a collinear momentum mode. This defines another perturbative off-shell mode that should be integrated out. The matching of QCD to the low-energy EFT thus has to be performed in two steps, 
    $$
      \mbox{QCD} \, \longrightarrow \, \mbox{SCET-I} \, \longrightarrow \, \mbox{SCET-II} \,,
    $$
    where in the first (second) step, the short-distance functions related to hard (hard-collinear) momentum modes are identified and calculated in renormalization-group improved perturbation theory.

\begin{figure}[t!p]
\begin{center}
\resizebox{0.95\columnwidth}{!}{%
\includegraphics{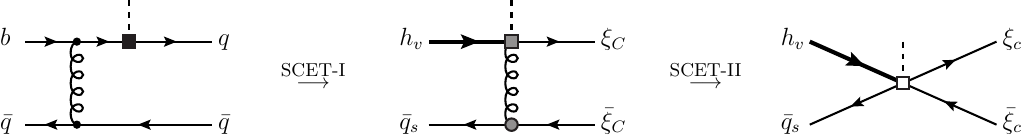} 
}

\end{center}
\caption{Example for the matching of a  heavy-to-light current from QCD to SCET, where the momentum transfer to the spectator quark is induced by a hard-collinear transverse gluon.}
\label{fig:diagrams_FFmatching}  
\end{figure}

    An example is shown in Fig.~\ref{fig:diagrams_FFmatching}, which corresponds to a tree-level diagram contributing to the so-called factorizable part of heavy-to-light form factors in the QCD factorization approach, where in the first step one integrates out the hard momentum fluctuations of the heavy $b$-quark, and in the second step the hard-collinear momentum modes of a transverse gluon  coupled to the spectator quark.

     In this context, it is sometimes convenient to formulate the SCET Lagrangian and currents in terms of manifestly gauge-invariant building 
     blocks,\footnote{For a recent discussion of the advantages of this formulation  for describing factorization at sub-leading order in the 
     power expansion, see also~\cite{Boer:2023yde}.}
     see e.g.~\cite{Hill:2002vw}. 
     For instance, in SCET-I
     one defines building blocks for hard-collinear quarks and hard-collinear transverse gluons via 
     \begin{align}
        \chi_C(x) & = W_C^\dagger(x) \, \xi_C(x) \,, \qquad
        {\cal G}_\perp^\mu(x) = W_C^\dagger(x) \, iD_\perp^\mu  W_C(x) - i \partial_\perp^\mu \,.
     \end{align} 
     Here, 
     the relevant spinor components for the hard-collinear quark field moving in the $n_-$ direction with $p_C^\mu \simeq E \, n_-^\mu$, are given by
     \begin{eqnarray}
     \xi_C = \frac{\slashed n_- \slashed n_+}{4} \, q_C  \,, \qquad n_\pm^2=0 \,, \quad n_- \cdot n_+ = 2 \,. 
     \end{eqnarray}
     The associated hard-collinear QCD Wilson lines are defined as the path-ordered exponential
     \begin{align}
         W_C(x) &= {\sf P} \, \exp\left(i g_s \, \int_{-\infty}^0 ds \, n_+ \cdot G_C(x+sn_+) \right) \,, 
         \qquad (i n_+ \cdot D) \, W_C(x) = 0 \,,
     \end{align}
     where $G_C^\mu(x)$ is the chromomagnetic 4-vector potential for the hard-collinear gluons.

     For instance, for the example shown in Fig.~\ref{fig:diagrams_FFmatching}, integrating out the hard $b$-quark propagator in the Feynman diagram on the right leads to effective decay currents in SCET-I,
     describing the decay of a heavy quark into a hard-collinear quark and a hard-collinear transverse gluon -- depicted by the shaded box. 
     In terms of hard-collinear building blocks and the HQET field they take the form \begin{align}
         J_i^{(1)}(x) &= \bar \chi_C(x) \, \slashed {\cal G}_\perp(x+sn_+) \, \Gamma_i \, h_v(x_-) \,,
         \label{eq:J1}
     \end{align}    
     with some Dirac matrix $\Gamma_i$,
     where the multipole expansion of the soft fields is accounted for by the short-hand notation 
     $x_-^\mu = \frac{n_+ \cdot x}{2} \, n_-^\mu$.
     In SCET-I the hard-collinear gluons from the decay vertex connect to the would-be spectator quark by a soft-collinear interaction term in the SCET-I Lagrangian that describes the effective vertex depicted as a shaded circle at the lower quark line in Fig.~\ref{fig:diagrams_FFmatching}.
     In terms of the collinear building blocks it can be written as 
     \begin{align}
         {\cal L}_{\xi q}^{(1)}(x) &= \bar q_s(x_-) \, \slashed {\cal G}_\perp(x) \, \chi_C(x) + \mbox{h.c.} 
         \label{eq:Lxiq1}
     \end{align}

 Similar building blocks can be defined for the SCET-II fields (denoted by $\chi_c$) that appear after integrating out the hard-collinear fluctuations, which for the example in Fig.~\ref{fig:diagrams_FFmatching}  are given as effective non-local 4-quark operators with soft and collinear fields (depicted by the white box). Specific examples will be discussed 
 in Sec.~\ref{sec:lcda} below.

\subsection{Renormalization-group factors in SCET}

As usual, operators in the effective theory have to be renormalized, leading to a non-trivial scale evolution of operator matrix elements and the associated short-distance coefficients. 
The solution of the RGEs
for SCET operators takes a generic form which can be expressed in terms of perturbatively calculable RG functions,\footnote{In the literature one also finds the notation $g(\mu,\mu_0)=-a(\mu,\mu_0)$.}
\begin{eqnarray}
  V(\mu,\mu_0) &=& - \int\limits_{\alpha_s(\mu_0)}^{\alpha_s(\mu)} \frac{d\alpha}{\beta(\alpha)} \left( \gamma(\alpha) + \Gamma_c(\alpha) \, \int\limits_{\alpha_s(\mu_0)}^\alpha \frac{d\alpha'}{\beta(\alpha')} \right) \,,\nonumber \\[0.2em]
  a(\mu,\mu_0) &=& - \int\limits_{\alpha_s(\mu_0)}^{\alpha_s(\mu)} \frac{d\alpha}{\beta(\alpha)} \, \Gamma_c(\alpha) \,.
\end{eqnarray}
Here $\beta(\alpha_s)$ is the QCD $\beta$-function and $\gamma(\alpha_s)$ an operator-specific anomalous dimension. 
The universal cusp anomalous dimension $\Gamma_c(\alpha_s)$ appears as a consequence of Wilson lines in the different light-like (or time-like)
directions.

In the leading-logarithmic  (LL) approximation -- where we restrict ourselves to terms of order $1/\alpha_s$ in the exponent -- the integrals in the RG function can be performed explicitly, and the result can be written
as an RG factor for the QCD-evolution effects
from  $\mu_0\to \mu_1$
\begin{eqnarray}
&&U_\omega(\mu_1,\mu_0;\mu_*)\Big|_{\rm LL} 
\nonumber \\[0.2em] &=& 
\exp\left[ \frac{4\pi C_F}{\beta_0^2} \left( \frac{1}{\alpha_s(\mu_0)} - \frac{1}{\alpha_s(\mu_1)} 
 + \frac{1}{\alpha_s(\mu_*)} \, \ln \frac{\alpha_s(\mu_0)}{\alpha_s(\mu_1)} \right) + \frac{2C_F}{\beta_0} \, \ln \frac{\alpha_s(\mu_0)}{\alpha_s(\mu_1)} \, \ln \frac{\omega}{\mu_*} \right] \,. 
 \cr &&
 \label{eq:Uw}
 \end{eqnarray}
Here $\beta_0$ is the first coefficient in the QCD $\beta$-function, and -- by introducing an auxiliary scale $\mu_*$ -- 
we have adopted a form where the group composition rule 
\begin{eqnarray}
U_\omega(\mu_2,\mu_1;\mu_*) \, U_\omega(\mu_1,\mu_0;\mu_*) = U_\omega(\mu_2,\mu_0;\mu_*)
\end{eqnarray}
is manifest (see also \cite{Feldmann:2014ika}). 
In concrete applications (see below) one identifies the auxiliary scale $\mu_*$ with the initial scale of the RG evolution,
defining
\begin{eqnarray}
U_\omega(\mu_1,\mu_0) \equiv  U_\omega(\mu_1,\mu_0;\mu_0) \,.
\end{eqnarray}

\subsection{Hadronic matrix elements in SCET-II}

\label{sec:lcda}

After factorizing the hard and hard-collinear dynamics into appropriate coefficient functions,
one is left with a number of universal hadronic input functions that are defined as matrix elements of non-local operators which involve 
soft or collinear field operators in SCET-II.
In this way, the energetic transitions probe light-like correlations inside the hadronic bound states which define light-cone distribution amplitudes (LCDAs) of light and heavy hadrons. 
Focusing on mesonic transitions at leading order in the heavy-quark expansion,
the relevant information is contained in matrix elements of 
two-body light-ray operators that are associated to the leading 2-particle Fock states in the initial and final-state mesons.

\subsubsection{Pion light-cone distribution amplitude}
For instance, the leading-twist LCDA for a charged pion with large momentum component $(n_+\cdot p)\simeq 2E$ in the $B$-meson rest frame is defined as 
\begin{equation}
\label{eq:intro:pionLCDA}
 \langle \pi^-(p)| \bar{\chi}^{(d)}(tn_+) \, \slashed n_+ \gamma_5 \, \chi^{(u)}(0)| 0\rangle  =
- i f_\pi \, (n_+ \cdot p) \, \int_0^1 du\, e^{i u (n_+ \cdot p) t} \, \phi_\pi(u,\mu) \,.
\end{equation}
Here the Wilson lines from the composite collinear-quark fields $\bar{\chi}^{(d)},\chi^{(u)}$ combine to a finite-length collinear Wilson line, reproducing the standard gauge link $[tn_+,0]$ between the two quark fields $u,d$. 
Further, $f_\pi$ is the pion decay constant in QCD, and $u$ corresponds to the 
light-cone momentum fraction of one of the quarks entering the hard-scattering process.
The so-defined pion LCDA obeys an RGE 
\begin{equation} \label{eq:pionDA_RGE}
  \frac{d\phi_\pi(u,\mu)}{d\ln\mu} =  -
  \int_0^1 dv \, \Gamma(u,v;\mu) \, \phi_\pi(v,\mu)    \,,
\end{equation}
which at 1-loop is determined by the ERBL kernel \cite{Lepage:1979zb,Lepage:1980fj,Efremov:1979qk},
\begin{equation}
\label{eq:ERBL}
    \Gamma(u,v;\mu) = - \frac{\alpha_s C_F}{\pi} \, V(u,v) + {\cal O}(\alpha_s^2) \,,
\end{equation}
with
\begin{align}  
V(u,v) = \bigg[ &\left(1+\frac{1}{v-u}\right) \frac{u}{v} \, \theta(v-u) + \left(1+\frac{1}{u-v}\right) \frac{\bar{u}}{\bar{v}}  \, \theta(u-v) \bigg]_+ \nonumber \\[0.2em]
+ &\left(u \ln \bar{u} + \bar{u} \ln u + \frac32 \right) \delta(u-v) \,.
\end{align}
Here we denote $\bar u = 1-u$, and $[\ldots ]_+$ refers to the standard plus-distribution.
Expanding the pion LCDA in Gegenbauer polynomials $C_n^{(3/2)}(x)$,
\begin{equation}
\label{eq:Gegenbauer1}
    \phi_\pi(u;\mu) = 6u \bar{u} \, \sum_{n=0}^\infty a_n(\mu) \, C_n^{\left( 3/2 \right)}(2u-1) \,,
\end{equation}
the Gegenbauer coefficients $a_n$ renormalize multiplicatively at 1-loop
\begin{equation}
    a_n(\mu) = \left( \frac{\alpha_s(\mu)}{\alpha_s(\mu_0)} \right)^{\frac{C_F \gamma_n}{\beta_0}} \, a_n(\mu_0) \,,
\end{equation}
with the anomalous dimension~\cite{Lepage:1979zb}
\begin{align}
    \gamma_n = 1-\frac{2}{(n+1)(n+2)} +4\sum_{m=2}^{n+1} \frac{1}{m} \,.
\end{align}

\subsubsection{$B$-meson light-cone distribution amplitude}
Similarly, for an exclusive $B$-meson decay with a large momentum transfer in the $n_-$ direction, the leading LCDA for a heavy $B$-meson in HQET is defined as 
\begin{align}
\label{eq:intro:BLCDA}
 &\langle 0| 
\bar{q}_s(tn_-) [tn_-,0] 
\,\slashed{n}_-\gamma_5 \, h_v(0) |\bar{B}_v\rangle 
= \, i F_{\rm stat}(\mu) \, \int_0^\infty d\omega\; e^{-i \omega t} \,\phi_B^+(\omega;\mu) \, .
\end{align}
Here, $F_{\rm stat}(\mu)$ is the decay constant in HQET, and $\omega$ corresponds to the light-cone momentum carried by the light-quark in the $B$-meson.
Its scale dependence 
\begin{equation}
  \frac{d\phi_B^+(\omega,\mu)}{d\ln\mu} =  -
  \int_0^\infty d\omega' \, \Gamma_B^+(\omega,\omega';\mu) \, \phi_B^+(\omega',\mu)    
\end{equation}
at one-loop is described by the Lange-Neubert kernel \cite{Lange:2003ff},
\begin{align} \label{eq:LNkernel}
    \Gamma_B^+(\omega,\omega';\mu) = \frac{\alpha_s C_F}{\pi} \bigg\{ &\left(\ln \frac{\mu}{\omega} - \frac12 \right)\delta(\omega-\omega') \nonumber \\[0.2em]
    - \omega&\left[ \frac{ ~\theta(\omega'-\omega)}{\omega'(\omega'-\omega)} + \frac{\theta(\omega-\omega')}{\omega(\omega-\omega')}\right]_+ \bigg\} + {\cal O}(\alpha_s^2) \,.
\end{align}
A particularly important feature of the $B$-meson LCDA that enters almost all QCD factorization theorems for exclusive decays
is captured by its inverse moment,
\begin{align}
  \lambda_B^{-1}(\mu) &\equiv \int_0^\infty \frac{d\omega}{\omega} \, \phi_B^+(\omega,\mu) \,.
\end{align}
The 1-loop evolution of this quantity can be written in compact form,
\begin{align}
 \lambda_B^{-1}(\mu) &= e^{V} \, \frac{\Gamma(1+a)}{\Gamma(1-a)} \,
 \int_0^\infty \frac{d\omega}{\omega} \left( \frac{\mu_0 \, e^{2\gamma_E}}{\omega} \right)^{\!a} \phi_B^+(\omega;\mu_0) \,,
 \label{eq:invmoment_evol_QCD}
\end{align}
with the RG functions $V=V(\mu,\mu_0)$ and $a=a(\mu,\mu_0)$ as defined above.
The continuous set of eigenfunctions for the Lange-Neubert kernel~(\ref{eq:LNkernel}) 
is given in terms of Bessel functions and can be found in Refs.~\cite{Bell:2013tfa,Braun:2014owa}.
 

\section{QED factorization in exclusive \ensuremath{B}-meson decays}
\label{sec:QEDF}

The modern EFT and factorization methods that we have sketched in the previous section also
provide a systematic way to describe the (virtual) QED radiation that resolves the substructure of composite mesons. 
Due to the electrically charged external particles, QED factorization theorems are in general more complicated than in QCD, 
and feature various additional subtleties like IR-divergent hadronic matrix elements.
In this chapter we summarize how the QCD factorization program can be extended by generalizing the results from the previous section. We will first discuss a number of generic results, 
before we review the recent progress in studying the factorization of structure-dependent QED effects in specific applications. 
This includes the purely leptonic modes \mbox{$B_{s,d}\to\mu^+\mu^-$} and $B^-\to\mu^- \bar{\nu}_\mu$~\cite{Beneke:2017vpq,Beneke:2019slt,Cornella:2022ubo}, as well as semi-leptonic and two-body non-leptonic decays~\cite{Beneke:2020vnb,Beneke:2021jhp,Beneke:2021pkl,Beneke:2022msp}.

\subsection{Generic discussion of QED effects}
QED corrections to $B$-meson decays constitute a complicated multi-scale problem, involving various typical energy scales ranging from the electroweak scale of order 100~GeV down to an experimental soft-photon resolution scale, typically of order a few$\times$10~MeV. Consequently, their theoretical description involves a tower of effective theories valid in a given energy range. Table~\ref{tab:scales} summarizes the corresponding effects of QED at various energies, which are also briefly discussed in the following. We stress that in this article we focus on the treatment of structure-dependent QED corrections using SCET, and that a complete EFT description at extremely soft scales has not yet been worked out.

\begin{table}[t!!pbh]
\begin{center}
\caption{\label{tab:scales} Summary of photon-energy ranges and associated EFTs that are relevant to incorporate QED effects
in energetic $B$-meson decays. Here, the perturbative analysis is valid for renormalization scales above a scale $\mu_0 \sim 1$~GeV, where the strong coupling $\alpha_s(\mu_0) \ll 1$. On the other hand, the description of QED effects in terms of a hadronic EFT is restricted to very low energy scales below a cutoff $\mu_{\rm IR}\sim$ of the order of a few 100~MeV. Finally, photons with energies below the experimental cut $\Delta E$ are not resolved.
}
\renewcommand{\arraystretch}{1.5}
\begin{tabular}{c|c|p{0.5\textwidth}}
Energy range & EFT & QED effects 
\\
\hline \hline 
$M_W \to m_b$ & weak Hamiltonian & matching and running of \newline electroweak Wilson coefficients  \newline (structure-independent)
\\
\hline \hline  
$m_b \to \sqrt{m_b \Lambda}$ & {SCET-I} & integrating out hard photons \newline RG running from hard-collinear photons
\\
\hline 
$\sqrt{m_b \Lambda} \to \mu_0$ & {SCET-II} & integrating out hard-collinear photons \newline RG running of soft and collinear functions
\\ 
$\mu_0 \to \mu_{\rm IR}$ & & 
hadronic functions depend on external charges
\\
\hline \hline 
\multicolumn{3}{c}{$\downarrow$ \quad non-perturbative matching \quad $\downarrow$ } 
\\
\hline 
$\mu_{\rm IR} \to \Delta E$ & hadronic EFT & photon radiation off point-like hadrons \newline 
cancellation of real and virtual IR divergences
\\
\hline 
\end{tabular}
\end{center}
\end{table}

\paragraph{Real photon emissions:} 

For exclusive $B$-meson decays into a specific final state $X$ 
involving charged particles in the initial or final state, any IR-finite observable must account for real radiation of undetected photons with arbitrarily small energy (``ultrasoft photons''). 
The soft-photon-inclusive decay rate
\begin{eqnarray}
 \Gamma^{B\to X}(\Delta E) = \Gamma^{B \to X + X_{\rm us}} \Big|_{E_{X_{\rm us}} \leq \Delta E}
\end{eqnarray}
provides such an IR-safe definition. Here $X_{\rm us}$ consists of photons (and electron-positron pairs) with \emph{total} energy $E_{X_{\rm us}}$ -- measured in the $B$-meson rest frame -- smaller than the experimental resolution $\Delta E$. 
In the following, it is assumed that this energy cut is \emph{smaller} than the QCD scale associated to the hadronic binding, $\Delta E \ll \Lambda$.
In this limit, the soft-photon inclusive decay width factorizes into an ultrasoft factor ${\cal S}(\Delta E)$, accounting for the radiation of ultrasoft photons from charged leptons or hadrons, and the absolute-square of the so-called \emph{non-radiative amplitude} ${\cal A}(B \to X)$ of the process,
\begin{eqnarray}
 \Gamma^{B\to X}(\Delta E) \sim |{\cal A}(B \to X)|^2 {\cal S}(\Delta E) \,.
\end{eqnarray}
The non-radiative amplitude is IR-divergent (or, equivalently, dependent on an IR subtraction scale $\mu_{\rm IR}$, see Table~\ref{tab:scales}), 
and these divergences cancel only in the product with the (divergent) radiation factor $S(\Delta E)$. 
Large logarithms of the cut-off $\Delta E$ and the masses of the involved particles then exponentiate in the double-logarithmic approximation, leading to the standard Bloch-Nordsieck factors (see~\cite{Beneke:2020vnb} for a detailed discussion).
For the reminder of this review article, we focus on the calculation of structure-dependent QED effects in the non-radiative amplitude, which provide another source of large logarithms and go beyond this simple approximation.

\paragraph{QED effects in electroweak Wilson coefficients:}

Radiative QED corrections from the electroweak scale can be included in a conceptually straightforward way. They are part of the Wilson coefficients $C_i(\mu)$ and have been calculated e.g. in~\cite{Gambino:2001au,Bobeth:2003at}. RG evolution from the electroweak scale down to the $b$-quark mass then incorporates large logarithms $\ln(M_W^2/m_b^2)$. For phenomenological applications it is usually sufficient to sum such logarithms in QCD, while performing a fixed-order expansion in the electromagnetic coupling $\alpha$, cf. e.g.~\cite{Huber:2005ig}. It is worth noting that, in contrast to QCD, also charged-current semi-leptonic $b \to u \ell^- \bar{\nu}_\ell$ transitions receive short-distance corrections in QED~\cite{Sirlin:1981ie}.

\paragraph{Gauge-invariant building blocks:}

In the presence of electromagnetic interactions in SCET, we also encounter QED Wilson lines associated to charged fermion fields.
For instance, for hard-collinear fermions $f$ with charge factor $Q_f$ these are defined as 
     \begin{align}
         W^{(Q_f)}_{C}(x) &= \exp\left(i e Q_f \, \int_{-\infty}^0 ds \, n_+ \cdot A_C(x+sn_+) \right) \,,
     \end{align}
     where $A_C^\mu(x)$ is now the electromagnetic 4-vector potential for hard-collinear photons.
With this we can modify the SCET building blocks to account for electromagnetic gauge invariance in the SCET Lagrangian and currents,
\begin{align} \label{eq:QEDquarkblocks}
        \chi^{(q)}_C(x) & = \left[W^{(Q_q)}_{C} W_C \right]^\dagger(x) \, \xi^{(q)}_C(x) &&\mbox{for hard-collinear quarks} \,,
        \cr 
        \ell_C(x) &= W^{(Q_\ell) \dagger}_{C}(x) \, l_C(x) && \mbox{for hard-collinear leptons}  \,.
\end{align}

\paragraph{QED evolution of SCET operators:}
Similarly as in Eq.~(\ref{eq:Uw}),
QED radiative corrections contribute to the RG equations for SCET operators. 
At LL accuracy, these take the generic form
\begin{eqnarray} 
&& U_\omega^{\rm em}(\mu_1,\mu_0;\mu_*; P_Q) \Big|_{\rm LL}
\cr 
&=&
 \exp\left[ \frac{4\pi \, P_Q}{\beta_{0,\rm em}^2} \left( 
 \frac{1}{\alpha(\mu_0)} - \frac{1}{\alpha(\mu_1)} 
 + \frac{1}{\alpha(\mu_*)} \, \ln \frac{\alpha(\mu_0)}{\alpha(\mu_1)} \right)
  + \frac{2P_Q}{\beta_{0,\rm em}} \, \ln \frac{\alpha(\mu_0)}{\alpha(\mu_1)} \, \ln \frac{\omega}{\mu_*}
 \right] \,,
 \cr &&
\end{eqnarray}
with $\beta_{0,\rm em}$ being the first coefficient of the QED $\beta$-function. In contrast to the QCD case,
the RG function now depend on the charge factors of the involved leptons and quarks in terms of a process-dependent quadratic polynomial
$P_Q$.

\paragraph{Hadronic input functions:}
 In the presence of QED corrections, hadronic input functions like transition form factors, or the LCDAs for light and heavy mesons need to be generalized. 
 As they appear in the factorization of virtual QED corrections in the \emph{non-radiative} amplitude, a peculiar feature of these functions is that they are defined by IR divergent hadronic matrix elements.   
 Furthermore, soft photons do no longer decouple from electrically charged external particles, and the QED generalization of the $B$-meson LCDA becomes a \emph{process-specific} soft function, which in general is complex-valued and contains effects like rescattering phases.
 More details on the QED-generalized LCDAs will be discussed in a dedicated Section~\ref{subsec:QED_LCDAs}.

\paragraph{Radiation of ultrasoft photons from pointlike hadrons:}

At very low photon energies, a certain class of universal \emph{structure-independent} QED effects can be described by an effective theory that only contains (point-like) hadrons. Such a framework -- together with some additional simplifying assumptions -- has been used e.g.\ in \cite{Bordone:2016gaq,Isidori:2020acz,Baracchini:2005wp} to investigate QED corrections to rare $B \to  K \ell^+\ell^-$ or non-leptonic two-body $B \to M_1 M_2$ decays. How to systematically incorporate the structure-dependent effects that we will discuss below into such a low-energy hadronic EFT is yet an open question.

\subsection{Leptonic decays}
\label{subsec:leptonic}

The first systematic development of QED factorization in exclusive $B$ decays has been discussed for the rare leptonic decays $B_{q} \to \mu^+\mu^+$, with $q = s,d$,
in \cite{Beneke:2017vpq,Beneke:2019slt}. 
Within the SM, these are induced by the weak effective Hamiltonian for $b \to s $ or $b\to d$ transitions, where in the following we focus on the $b \to s$ case. 
The most relevant operators in (\ref{eq:intro:Heff}) are the  semi-leptonic four-fermion operators 
\begin{eqnarray}
    {\cal O}_9 &=& \frac{\alpha}{4\pi} \left( \bar s \gamma^\mu P_L b\right) \left(\bar l \gamma_\mu l \right) \,, \qquad 
    {\cal O}_{10} = \frac{\alpha}{4\pi} \left(\bar s \gamma^\mu P_L b\right) \left( \bar l \gamma_\mu\gamma_5 l\right)  \,,
\end{eqnarray}
together with the electromagnetic penguin operator
\begin{eqnarray}
  {\cal O}_7 &=& \frac{e}{(4\pi)^2} \, \bar m_b \left( \bar s \sigma^{\mu\nu} P_R b \right) F^{\mu\nu} \,,
\end{eqnarray}
where $l(x)$ denotes the leptonic field operators (in this case for muons), and $P_{R,L} = \frac{1\pm \gamma_5}2$ are the standard chiral projectors.

In the absence of QED effects, only the operator ${\cal O}_{10}$ contributes, and the hadronic effects are contained in the matrix 
element of the \emph{local} $b \to s$ current,
\begin{eqnarray}
  \langle 0 |\bar s \gamma^\mu\gamma_5 b |\bar B_s(p)\rangle &\equiv & i f_{B_s} p^\mu \,,
\end{eqnarray}
which defines the $B_s$-meson decay constant $f_{B_s}$. 
Furthermore, as a consequence of helicity conservation,
the resulting decay amplitude in the SM is proportional to the muon mass (denoted by $m$ in the following) which results in a suppression factor compared 
to other $B$ decay modes. This makes $B_s \to \mu^+\mu^-$ a particularly interesting mode for indirect searches for 
new interactions beyond the SM, which, in turn, requires to understand sizable corrections \emph{within the SM}, as well.

\begin{figure}[t!p]
\begin{center}
\resizebox{0.9\columnwidth}{!}{%
\includegraphics{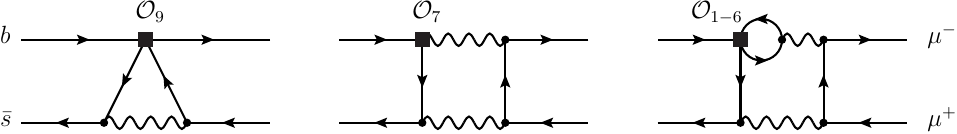} 
}

\end{center}
\caption{Example diagrams, leading to power-enhanced QED corrections in $\bar B_s \to \mu^+\mu^-$.
}
\label{fig:diagrams_Bsmumu}
\end{figure}

Including QED interactions in $B_s \to \mu^+\mu^-$ decays spoils the factorization of hadronic and leptonic currents. Relevant example diagrams are shown in Fig~\ref{fig:diagrams_Bsmumu}. 
This leads results in a number of interesting new features:
\begin{enumerate}
\item[(i)] \emph{Soft enhancement:}
The exchange of photons between the light spectator quark in the  $\bar B_s$-meson and the final-state muons can lift the helicity suppression, 
$$
  \frac{m}{m_B} \longrightarrow \frac{m}{\lambda_B} \sim {\cal O}(1)\,,
$$
where $\lambda_B  \sim {\cal O}(\Lambda)$ arises from the propagator of the resolved spectator quark and requires information on a \emph{non-local} hadronic 
matrix element which defines the (generalized) light-cone distribution amplitudes of the $B_s$ meson. 

\vskip0.25em

\item[(ii)] \emph{Enlarged operator basis:} In QED, also the semi-leptonic operator ${\cal O}_9$ and the electromagnetic penguin operator ${\cal O}_7^\gamma$ in the weak effective Hamiltonian contribute, together with universal 1-loop effects from 4-quark operators.

\vskip0.25em

\item[(iii)] \emph{Rapidity logarithms from endpoint divergences:} 
The contribution of the electromagnetic dipole operator ${\cal O}_7$ features endpoint-divergent convolution integrals in the standard factorization approach.
This leads to a double-logarithmic enhancement  
$\sim \ln^2 \frac{m^2}{m_{B}\lambda_B}$
with
$$\frac{m^2}{m_B \lambda_B} \sim \frac{\Lambda}{m_B} \equiv \lambda^2  \ll 1 \,, $$
where we have counted the muon mass $m$ as of the same order as the scale $\Lambda$ associated to soft hadronic binding effects. 

\end{enumerate}

Taking only the leading-order contribution of the power-enhanced QED corrections
into account, the authors of Ref.~\cite{Beneke:2017vpq} find the following form for the $B_s\to \mu^+\mu^-$ decay amplitude
\begin{eqnarray}
    i{\cal A} &=& i {\cal A}_{10} + i {\cal A}_9 + i {\cal A}_7 
    \nonumber \\[0.2em] &=& {\cal N} \, f_{B_s}  m \Bigg\{ 
    \left[ \bar \ell \gamma_5 \bar \ell \right]  C_{10}
    \nonumber \\[0.2em] 
    &&  \quad {} + \frac{\alpha}{4\pi} \, Q_\mu  Q_s \, 
     M \left[\bar \ell \, (1+\gamma_5) \, \ell \right] 
     \nonumber \\[0.2em]
     && \qquad {} \times \Big(
     \int_0^1 du \, (1-u) 
     \int_0^\infty \frac{d\omega}{\omega} \, \phi_B^+(\omega) \left( \ln \frac{m_B \omega}{m^2} + \ln \frac{u}{1-u} \right) C_9^{\rm eff}(um_b^2)
     \nonumber \\[0.2em] 
     && \qquad \quad {}
     - Q_\mu \, \int_0^\infty \frac{d\omega}{\omega} \, \phi_B^+(\omega) 
     \left( \ln^2 \frac{m_B \omega}{m^2} - 2 \ln \frac{m_B \omega}{m^2} + \frac{2\pi^2}{3} \right) \, C_7^{\rm eff} \Big) \Bigg\} + \ldots 
     \cr &&
     \label{eq:dimuon:amp}
\end{eqnarray}
where $u$ corresponds to the light-cone momentum fraction carried by the lepton in the loop diagrams in Fig.~\ref{fig:diagrams_Bsmumu}, 
see also Eq.~(\ref{eq:udefinition}) below,
and the common normalization factor has been defined as
\begin{eqnarray}
 {\cal N} = V_{tb} \, V_{ts}^* \, \frac{4G_F}{\sqrt 2} \, \frac{\alpha}{4\pi} \,.
\end{eqnarray}
In the following, we briefly summarize the main steps in the EFT construction which promote the above result to a QCD/QED factorization theorem for the contributions of 
 the operators ${\cal O}_9$ and ${\cal O}_{10}$.  The contribution from the electromagnetic penguin operator ${\cal O}_7$ turns out to be more subtle, and a dedicated discussion will be given in Section~\ref{sec:O7} further below.
 We also mention that a similar analysis has been performed for $B_s \to \tau^+\tau^-$ decays in \cite{Huang:2023nli}, where an analogous soft enhancement occurs. 
 However, in this case the $\tau$ mass is parametrically of order of the hard-collinear scale, and therefore the corresponding logarithms in Eq.~(\ref{eq:dimuon:amp}) are not large, and the overall numerical effect of QED corrections is very small.

\subsubsection{Matching to {SCET-I} and RG evolution}

In the rest frame of the decaying $B_s$ meson, the final-state muons are emitted back-to-back with energies $E \simeq m_B/2$. In the EFT approach these 
will therefore be described by collinear and anti-collinear lepton fields, recoiling against soft quarks and gluons in the presence of a quasi-static $b$-quark.

In the first step of the EFT construction, hard modes with virtualities $\mu^2 \sim m_b^2$ are integrated out. In the given case, only 
the following {SCET-I} operators turn out to be relevant \cite{Beneke:2019slt},
\begin{eqnarray}
 \tilde{\cal O}_9(s,t) &=& g_{\mu\nu}^\perp \left[\bar \chi^{(q)}_C(sn_+) \, \gamma_\perp^\mu P_L h_v(0) \right] \left[\bar \ell_C(tn_+) \, \gamma_\perp^\nu \, \ell_{\bar C}(0) \right] \,,
 \cr 
 \tilde{\cal O}_{10}(s,t) &=& \epsilon_{\mu\nu}^\perp \left[\bar \chi^{(q)}_C(sn_+) \, \gamma_\perp^\mu P_L h_v(0) \right] \left[\bar \ell_C(tn_+) \, \gamma_\perp^\nu \, \ell_{\bar C}(0) \right] \,,
\end{eqnarray}
where
\begin{eqnarray}
g_{\mu\nu}^\perp \equiv g_{\mu\nu} - \frac{n_+^\mu n_-^\nu}{2}- \frac{n_-^\mu n_+^\nu}{2} \quad \mbox{and} \quad \epsilon_{\mu\nu}^\perp= \epsilon_{\mu\nu\alpha\beta} \, \frac{n_+^\alpha n_-^\beta}{2} \,. 
\end{eqnarray}
By the symmetry of the setup, another set of operators, $\tilde{\cal O}_{\bar 9, \overline{10}}(s,t)$,
is obtained by interchanging the role of $n_+ \leftrightarrow n_-$,  and of hard-collinear and anti-hard-collinear fields, see \cite{Beneke:2019slt}.
After Fourier transformation, the matching coefficients for the operators $\tilde O_{9,10}$ will represent the hard functions in the factorization theorem. 
For the LL accuracy, it is sufficient to perform the
matching calculation from QCD/QED to {SCET-I} at tree-level. 
In this case the matching coefficients at a hard scale $\mu =\mu_h \sim m_b$ 
are simply proportional to the corresponding Wilson coefficients from the weak effective Hamiltonian,
\begin{eqnarray}
  H_9(u,\mu_h) = {\cal N} \, C_9^{\rm eff}(u s_{\ell\bar \ell},\mu_h) \,, \qquad 
  H_{10}(\mu_h) = {\cal N} \,  C_{10}(\mu_h) \,.
  \label{eq:dimuon:H9tree}
\end{eqnarray}
Here $u$ is the light-cone momentum fraction of the hard-collinear lepton w.r.t.\ the total hard-collinear momentum,
\begin{equation} \label{eq:udefinition}
u = \frac{n_+ \cdot p_\ell}{n_+\cdot p_\ell + n_+ \cdot p_\chi} \,.
\end{equation}
Here $C_9^{\rm eff}$ already 
contains the universal 1-loop contributions from the 4-quark operators, with $s_{\ell\bar \ell}$ being the invariant dimuon mass.
The scale-dependence of the hard functions is determined by the RGEs in {SCET-I}. At LL accuracy, the solution can be written as \cite{Beneke:2019slt}
\begin{eqnarray}
 H_i(u,\mu) &\stackrel{\rm LL}{=}& U_h(\mu,\mu_h) \, H_i(u,\mu_h) \,,
 \label{eq:dimuon:H9evol}
 \end{eqnarray}
 with 
 \begin{eqnarray}
     U_h(\mu,\mu_h) &=& U_{m_B}(\mu,\mu_h) \, U_{m_B}^{\rm em}(\mu,\mu_h;P_Q) \,, \qquad P_Q= 2Q_\ell^2+Q_s \, (2 Q_\ell + Q_s) \,,
\end{eqnarray}
as long as flavour thresholds in the interval $[\mu,\mu_h]$ are ignored. 

\subsubsection{Matching to {SCET-II}}

The second step in the EFT construction consists of integrating out hard-collinear modes with virtualities of order $\mu^2 \sim m_b \Lambda$. This amounts
to matching the above {SCET-I} operators to an operator basis in {SCET-II} which only contains soft and collinear momentum modes. Here the light quarks 
in the $B_s$ meson are treated as massless, while in the collinear Lagrangian for the lepton fields the muon mass is kept. 
The various splittings of the hard-collinear fields into soft and collinear degrees of freedom technically amounts to evaluating time-ordered products 
of the {SCET-I} operators with sub-leading terms in the {SCET-I} Lagrangian. These can be analyzed along the lines  
worked out in \cite{Beneke:2003pa} for the case of heavy-to-light form factors (see also~\cite{Boer:2023yde}). 
For the power-enhanced electromagnetic contributions at LL accuracy, one finds \cite{Beneke:2019slt} that only two types of operators are relevant,
\begin{eqnarray}
  \tilde {\cal J}_{m\chi}^{A1}(t) &=& 
  \left[\bar q_s(tn_-) \, Y^{(Q_q)}(tn_-,0) \, \frac{\slashed n_-}{2} \, P_L \, h_v(0) \right]
  \left[Y_+^{(Q_\ell) \dagger} Y^{(Q_\ell)}_-\right](0) 
  \cr  && \qquad {} \times
  \left[\bar \ell_c(0) \, (4m P_R) \, \ell_{\bar c}(0)\right] \,, 
  \nonumber \\[0.25em]
  \tilde {\cal J}_{A\chi}^{B1}(t,s) &=& \left[\bar q_s(tn_-) \, Y^{(Q_q)}(tn_-,0) \, \frac{\slashed n_-}{2} \, P_L \, h_v(0) \right]
  \left[Y_+^{(Q_\ell) \dagger} Y_-^{(Q_\ell)}\right](0)
  \cr  && \qquad {} \times 
  \left[\bar \ell_c(0) \, (2 {\cal A}_{c\perp}(sn_+) P_R) \, \ell_{\bar c}(0) \right] \,,
  \label{eq:leptonic-scet2}
\end{eqnarray}
and analogous operators with the role of $n_+$ and $n_-$ interchanged. Both operators have the same scaling in the expansion parameter $\lambda$ 
and mix under {SCET-II} renormalization.

In the above operators, the quark fields are delocalized along the $n_-$ light-cone, where the corresponding gauge-link is given by 
soft Wilson lines containing QCD and QED gauge fields,
\begin{eqnarray}
  Y^{(Q_q)}(x,y) &=& \exp\left[ i e Q_q \int_y^x dz_\mu \, A_s^\mu(z) \right] {\cal P} \exp\left[ i g_s \int_y^x dz_\mu G_s^\mu(z)\right] \,.
\end{eqnarray}
The product of QED Wilson lines $[Y_+^{(Q_\ell) \dagger} Y_-^{(Q_\ell)}]$ results from the decoupling of soft photons from the collinear and 
anti-collinear leptons, where 
\begin{eqnarray}
 Y^{(Q_\ell)}_\pm(x) &=& \exp\left[ -ieQ_\ell \, \int_0^\infty ds \, n_\mp \cdot A_s(x+s n_\mp)\right]
\end{eqnarray}
is a semi-infinite Wilson line extending to $+\infty$ for final-state muons.

Matching the {SCET-I} operators from above onto this set of {SCET-II} operators introduces the hard-collinear matching coefficients (aka jet functions).
After Fourier transformation to momentum space, the matching relation reads
\begin{eqnarray}
  {\cal O}_i(u) &=& \int d\omega \left[ J_m(u,\omega) \, {\cal J}_{m\chi}^{A1}(\omega)
  + \int dw \, J_A(u,\omega,w) \, {\cal J}_{A\chi}^{B1}(\omega,w) 
  \right ] \,.
\end{eqnarray}
At tree-level, one has 
\begin{eqnarray}
  J_A^{(0)}(u,\omega,w) = - \frac{Q_q}{\omega} \, \delta(u-w) \,,
  \label{eq:leptonic:JA}
\end{eqnarray}
wheres the jet function $J_m(u,\omega)$ only starts at 1-loop level, with 
\begin{eqnarray}
 J_m^{(1)}(u,\omega) &=& \frac{\alpha}{4\pi} \, Q_\ell Q_q \, \frac{1-u}{\omega} \left[ 
  \ln \left( \frac{2 E_\ell \, \omega}{\mu^2} \right) + \ln \left(u(1-u)\right)
 \right] \theta(u) \theta(1-u) \,.
 \label{eq:leptonic:Jm}
\end{eqnarray}

\subsubsection{Renormalization in {SCET-II}}

The relevant {SCET-II} operators above factorize into a soft, a collinear and an anti-collinear sector. The {SCET-II} Lagrangian does not contain interactions between the soft and (anti-)collinear modes anymore, and therefore one would naively expect that the renormalization can be performed for the soft, collinear and anti-collinear sector independently.
However, it turns out that there is a factorization anomaly which is related to the overlap of soft and (anti-)collinear regions in the contributing loop diagrams.
For instance, at 1-loop order one explicitly finds an IR singularity arising from a photon-tadpole diagram 
which is only canceled by 
corresponding singularities in the collinear and anti-collinear sector. 
As the renormalization constants in the individual sectors must not depend on the IR regulator, one has to define a suitable subtraction scheme. 
One possibility \cite{Beneke:2003pa} is to factor out the vacuum expectation value of the product of soft QED Wilson lines, 
\begin{eqnarray}
   \tilde {\cal J}_s(t) & \equiv & \frac{\left[\bar q_s(tn_-) \, Y^{(Q_q)}(tn_-,0) \, \frac{\slashed n_-}{2} \, P_L \, h_v(0) \right]
  \left[Y_+^{(Q_\ell) \dagger}(0) Y_-^{(Q_\ell)}(0)\right] }{\langle 0|Y_+^{(Q_\ell) \dagger}(0) Y_-^{(Q_\ell)}(0)|0\rangle} \,.
\end{eqnarray}
The perturbative expansion of the denominator cancels the tadpole diagrams with soft photons, while the 
remaining renormalization constants will only depend on the total electric charge in the collinear and anti-collinear direction.
After Fourier transformation, the soft operator ${\cal J}_s(\omega,\mu)$ will depend on
the light-cone momentum component $\omega$ of the light quark. 

The corresponding renormalized matrix element with the $B_s$-meson state defines
a soft function which can be considered a generalization of the leading light-cone distribution amplitude $\phi_B^+(\omega)$, see Eq.~(\ref{eq:QED_BLCDA}). A dedicated discussion of these functions and their renormalization, also for the case of two-body decays of charged $B$ mesons, will be given in Sec.~\ref{sec:QED-B-LCDA} further below.

At LL accuracy 
the evolution of the soft operator can again be described 
by a multiplicative RG factor, such that 
\begin{eqnarray}
  {\cal J}_s(\omega;\mu) &\stackrel{\rm LL}{=} & U_s(\mu,\mu_s;\omega) \, {\cal J}_s(\omega;\mu_s) \,,
\end{eqnarray}
with 
\begin{eqnarray}
  U_s(\mu,\mu_s;\omega) &=& U_\omega(\mu,\mu_s) \, U_\omega^{\rm em}(\mu,\mu_s;P_Q) \,, \qquad P_Q =Q_s \, (2 Q_\ell +Q_s) \,.
\end{eqnarray}
Here $\mu_s$ is a soft scale, satisfying $\omega \sim {\cal O}(\mu_s) \sim \Lambda$.

In order to renormalize the collinear and anti-collinear sectors of the {SCET-II} operators in Eq.~(\ref{eq:leptonic-scet2}), one
has to divide the subtraction factor  -- which has been introduced above to cancel the IR divergences in the renormalization for the soft sector --
into two factors,
\begin{eqnarray} \label{eq:Rplusminus}
  \langle 0|Y_+^{(Q_\ell) \dagger}(0) Y_-^{(Q_\ell)}(0)|0\rangle & \equiv & R_+ R_- \,.
\end{eqnarray}
It is to be stressed that this decomposition is not unique, as one can always redistribute \emph{IR-finite} terms between $R_+$ and $R_-$
which thus corresponds to different renormalization schemes, see also footnote~\ref{footnote:conventions}.
The choice pursued in \cite{Beneke:2019slt} is a symmetric one, where $R_+ \leftrightarrow R_-$ under exchange of $n_+ \leftrightarrow n_-$.

One can then proceed to renormalize the anti-collinear sector, which for both {SCET-II} operators in Eq.~(\ref{eq:leptonic-scet2}) 
consists of a single anti-collinear lepton field, such that the collinear operator to be renormalized is defined as
\begin{eqnarray}
  {\cal J}_{\bar c} &=& R_- \, \ell_{\bar c}(0) \,,
  \label{eq:Rminus}
\end{eqnarray}
where the factor $R_-$ stems from the soft rearrangement. The operator matrix element with the anti-collinear lepton state is defined as
\begin{align} \label{eq:Zl}
    \langle \ell^+(p_{\bar \ell})| R_- \, \ell_{\bar c}(0)|0\rangle &= 
    Z_{\bar \ell} \, v(p_{\bar \ell}) \,,
\end{align}
with $Z_{\bar \ell}=1+{\cal O}(\alpha)$.
Again, at LL accuracy the solution of the resulting RG equations 
can be accounted for by a multiplicative factor,
\begin{eqnarray}
  {\cal J}_{\bar c}(\mu) &=& U_{\bar c}(\mu,\mu_c) \, {\cal J}_{\bar c}(\mu_c) \,, 
\end{eqnarray}
with 
\begin{eqnarray}
  U_{\bar c}(\mu,\mu_c) &=& 
  U_{m_B}^{\rm em}(\mu,\mu_c; P_Q) 
\,, \qquad P_Q = -Q_\ell^2 \,.
\end{eqnarray}

Finally, for the collinear sector of Eq.~(\ref{eq:leptonic-scet2}) one has to take into account the mixing between the two different collinear operators, 
\begin{eqnarray}
    {\cal J}_c^{\rm A1} &=& R_+ \, \bar \ell_c(0) \, 4m P_R \,, 
    \nonumber \\[0.25em]
    {\cal J}_c^{\rm B1} &=& R_+ \, (n_+ \cdot p) \, \int \frac{ds}{2\pi} \, e^{-i\overline{w} s \, (n_+ \cdot p)} \, \bar \ell_c(0) \, 2 \slashed {\cal A}_{c\perp}(sn_+) \, P_R \,,
\end{eqnarray}
with $\overline{w}=1-w$  being the light-cone momentum associated to the collinear photon field. The factor $R_+$ again ensures that the UV-divergences do not 
depend on the IR regulator. 
The matrix elements of these operators are given by
\begin{eqnarray}
  \langle \ell^-(p_\ell)|R_+ \bar \ell_c(0)|0\rangle &=& Z_\ell \, \bar u(p_\ell) \,, 
  \cr 
  R_+ \, \int \frac{ds}{2\pi} \, e^{-i\overline{w} s \, (n_+ \cdot p_\ell)} \, 
  \langle \bar \ell^-(p_\ell)|\bar \ell_c(0) \, {\cal A}_{c\perp}^\mu(sn_+)|0\rangle 
  &=& Z_\ell \, M_A(w) \, m_\ell \, [\bar u(p_\ell)\, \gamma_\perp^\mu] \,, 
\end{eqnarray}
which defines the function $M_A(w)$ as the Fourier transform of the light-cone separation between the collinear lepton and photon field, with the LO result
\begin{eqnarray}
  M_A(w) &=& - \frac{\alpha}{4\pi} \, Q_\ell \, \overline{w} \left( \ln \frac{\mu^2}{m_\ell^2} - \ln \overline{w}^2 \right) + {\cal O}(\alpha^2) \,.
  \label{eq:leptonic:MA}
\end{eqnarray}
At LL accuracy, the evolution of the collinear operators is given by 
\begin{eqnarray}
  {\cal J}_c^{A1}(\mu) &=& U_c(\mu,\mu_c) \, {\cal J}_{c}^{A1}(\mu_c) \,, 
  \nonumber \\[0.2em]
  {\cal J}_c^{B1}(\mu) &=& U_c(\mu,\mu_c) \left[ {\cal J}_c^{B1}(w;\mu_c) - \frac{Q_\ell \,\overline{w}}{\beta_{0,\rm em}} \, 
  \ln \frac{\alpha(\mu_c)}{\alpha(\mu)} \, {\cal J}_{c}^{A1}(\mu_c) \right] ,
\end{eqnarray}
with $U_c(\mu,\mu_c)=U_{\bar c}(\mu,\mu_c)$ defined above.
It is to be noticed that the contribution of the operator ${\cal J}_c^{B1}$ to $B_s \to \mu^+\mu^-$ only starts at 1-loop level, without logarithmic
enhancement when evaluated at the collinear scale $\mu_c$. One can therefore set ${\cal J}_c^{B1}(w;\mu_c)$ to zero at LL accuracy.

The above results can be summarized in terms of a proper
factorization theorem for the power-enhanced contribution to the $B_s \to \mu^+\mu^-$   decay amplitude,
\begin{eqnarray}
i{\cal A}_9 &=& T_+ \, \int_0^1 du \, 2 H_9(u) \, \int_0^\infty d\omega 
\cr && \qquad {} \times 
\left[J_m(u,\omega) + \int_0^1 dw \, J_A(u,\omega,w) \, M_A(w)] \right] m_{B_s} \, {\cal F}_{B_s} \Phi_B(\omega) \,,
\end{eqnarray}
where the common leptonic pre-factor has been defined as 
\begin{eqnarray}
  T_+(\mu) &\equiv & (-i) \, m_\ell(\mu) \, Z_\ell(\mu) \, Z_{\bar\ell}(\mu) \, 
  [\bar u(p_\ell) \, P_R \, v(p_{\bar \ell})] \,.
\end{eqnarray}
The tree-level matching for the hard function $H_9(u)$ is given in Eq.~(\ref{eq:dimuon:H9tree}), while 
the leading expressions for the jet functions $J_m(u,\omega)$ and $J_A(u,\omega,w)$
are given in Eqs.~(\ref{eq:leptonic:JA},\ref{eq:leptonic:Jm}).
The leading term for the function $M_A(w)$ can be found in Eq.~(\ref{eq:leptonic:MA}).
Finally, the hadronic matrix element of the soft operator defines a soft function $\Phi_B(\omega)$ which generalizes the leading $B$-meson LCDA, \begin{align}
    \langle 0| {\cal J}_s(t) |\bar B_s(p)\rangle 
    &= - \frac{i m_{B_s}}{4} \, \int_0^\infty d\omega \, e^{-i\omega t} \, {\cal F}_{B_s}  \Phi_B(\omega) \,,
\end{align}
together with a generalization of the $B_s$-meson decay constant ${\cal F}_{B_s}$ in QCD.
The properties of the so-defined soft function will be further discussed in Sec.~\ref{sec:QED-B-LCDA}, however in another scheme, see again footnote~\ref{footnote:conventions}.

The scale dependence of all functions is controlled by the RGEs in SCET as discussed above. In this way large logarithms of ratios of hard, hard-collinear and soft scales are resummed into the RG functions, improving the fixed-order result in Eq.~(\ref{eq:dimuon:amp}).

Notice that there is no power-enhanced contribution from the operator ${\cal O}_{10}$ in the final result, while the
factorization for the contribution of the operator ${\cal O}_7$ is more subtle, as already mentioned, and has been dropped here (but will be discussed separately below).

\subsection{Semi-leptonic and non-leptonic decays}
\label{subsec:nonleptonic}

\begin{figure}[t!p]
\begin{center}
\resizebox{0.95\columnwidth}{!}{%
\includegraphics{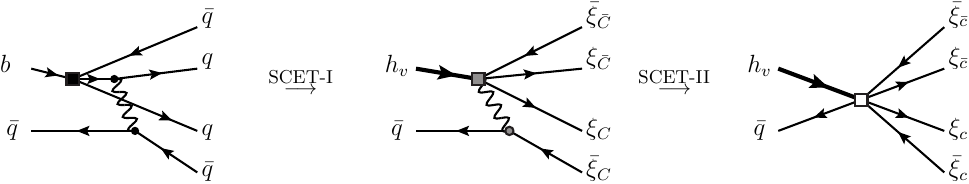} 
}

\end{center}
\caption{Example for an electromagnetic contribution to the matching of a  4-quark operator in QCD/QED to a 6-quark operator in SCET-II, as it appears in the factorization theorem for 
non-leptonic $B$-meson decays. Similar to the form factor case, see Fig.~\ref{fig:diagrams_FFmatching}, the photon is considered to be transversely polarized.}
\label{fig:diagrams_nonlep_matching}       
\end{figure}

The factorization of structure-dependent QED corrections to the non-leptonic decays $B_q \to M_1 M_2$ (with $q = u,d,s$ and two mesons $M_i$) 
has been analyzed in a series of papers~\cite{Beneke:2020vnb,Beneke:2021jhp,Beneke:2021pkl,Beneke:2022msp}. The final-state hadrons further complicate the analysis compared to the leptonic channels discussed in the previous section, and require introducing QED-generalized LCDAs for energetic light mesons as well as generalizations of hadronic transition form factors. 
A key result of the analyses is that QED corrections to the hadronic matrix elements preserve the structure of the well-established QCD factorization formulas, while the individual ingredients
need to be generalized, which is conceptually intricate in case of electrically charged mesons.

For concreteness, consider $B_q$ decays into two energetic pseudo-scalar mesons induced by the current-current operators 
\begin{align}
\label{eq:Q12}
 {\cal O}_1 &= [\bar{u} \gamma^\mu T^a (1-\gamma_5)b] \, [\bar{D} \gamma_\mu T^a (1-\gamma_5) u] \,, \nonumber \\
 {\cal O}_2 &= [\bar{u} \gamma^\mu (1-\gamma_5)b] \, [\bar{D} \gamma_\mu (1-\gamma_5) u] \,,
\end{align}
in the weak effective Hamiltonian,
where $\bar D=\{\bar d,\bar s \}$ is a light down-type quark.
In the heavy-quark limit,
the associated hadronic matrix elements can be expressed by a factorization formula
\begin{eqnarray}
\label{eq:non-leptonics-LL}
\left\langle M_1 M_2 | Q_i |\bar B_q \right\rangle &=& i m_B^2\,
\bigg\{
f_{M_2}\mathcal{F}^{BM_1}_{Q_2}(m_{M_2}^2) \, \int\limits_0^1 du \, 
T^{{\rm I}}_{i,Q_{2}}(u)  \, \Phi_{M_2}(u) 
\nonumber \\
&&\hspace*{-2.5cm} {} + f_{M_1}  f_{M_2}  f_{B} 
\,\int\limits_{-\infty}^\infty d \omega \int\limits_0^1 du \int\limits_0^1 dv \,
T^{{\rm II}}_{i, \otimes}(u,v,\omega) \,
\Phi_{M_1}(v) \, \Phi_{M_2}(u) \, \Phi_{B,\otimes} (\omega)
\,\bigg\} \,,
\end{eqnarray}
where by convention the meson $M_1$ picks up the spectator quark of the $B_q$ meson. 
The symbol $\otimes = (Q_1, Q_2)$ is introduced in Eq.~(\ref{eq:non-leptonics-LL}) as a short-hand notation to indicate the dependence on the electric charges of the two mesons $M_1$ and $M_2$, respectively. 

In the second term in~\eqref{eq:non-leptonics-LL}, typically called the ``hard-spectator-scattering term'', all mesons factorize and it can thus be expressed as a convolution of the three light-cone distribution amplitudes with a perturbative coefficient function $T^{\rm II}$.  
Both, the LCDAs as well as the hard-scattering kernel need to be generalized to account for virtual QED corrections. 
As an example, Fig.~\ref{fig:diagrams_nonlep_matching} shows the exchange of a hard-collinear photon yielding an $\mathcal{O}(\alpha)$ correction to $T^{\rm II}$.
It is to be stressed that the non-decoupling of soft photons from electrically charged final state particles results in a process dependence of the QED-generalization of the $B$-meson LCDA, which hence also carries the subscript $\otimes$.
For the case $\otimes = (+,-)$, this soft function agrees with the one that appears in $B_{s,d} \to \mu^+ \mu^-$, but in non-leptonic decays all charge combinations $\otimes = (0,0),(-,0),(0,-),(+,-)$ need to be considered.\footnote{\label{footnote:conventions} We note different conventions regarding the definitions of the hadronic functions that have been employed in the literature. First, in~\cite{Beneke:2020vnb,Beneke:2021jhp,Beneke:2021pkl,Beneke:2022msp} the rearrangement factors $R_c$ and $R_{\bar c}$ have been defined via a corresponding formula~\eqref{eq:Rplusminus}, but using the absolute value on the left-hand side. This definition avoids spurious imaginary parts in the collinear functions. Second, instead of introducing generalized decay constants, all QED effects have been absorbed in generalized LCDAs $\Phi_M$, $\Phi_{B,\otimes}$ in~\cite{Beneke:2021jhp,Beneke:2021pkl,Beneke:2022msp}. Decay constants coincide with their definition \emph{in absence} of QED in this scheme.} More details on the renormalization of the different soft functions will be discussed in Sec.~\ref{sec:QED-B-LCDA}.  

Contrarily, in the ``form-factor term'' -- the first term in~\eqref{eq:non-leptonics-LL} -- only the meson $M_2$ is decoupled from the $B \to M_1$ transition, and the respective LCDA $\Phi_{M_2}(u)$ is convoluted with a hard-scattering kernel $T^{\rm I}$. 
The quantity $\mathcal{F}^{BM_1}_{Q_2}$ should be identified with a $B \to M_1$ transition form factor at maximum recoil $q^2 = m_{M_2}^2 \simeq 0$ in QCD, and in QED if $M_2$ is uncharged.
However, soft-photon exchanges with an electrically charged meson $M_2$ undermine the QED analogue of the color-transparency argument 
~\cite{Beneke:1999br}.
In this case, soft photons again ``feel'' the presence of a charge moving in the $n_+$ direction, which therefore also requires to introduce a process-specific generalization of the form factor.

This generalization is related to the (non-radiative) semi-leptonic $\bar{B} \to M_1 \ell^- \bar{\nu}_\ell$ decay amplitude in the kinematic limit of a soft neutrino, whose factorization has also been studied in~\cite{Beneke:2020vnb}. 
More precisely,
\begin{equation} \label{eq:FAred}
    \mathcal{F}^{BM_1}_{-} = \frac{{\cal A}^{{\rm sl},M_1}_{\rm red}}{C_{\rm sl} Z_\ell} \,,
\end{equation}
with the reduced semi-leptonic amplitude ${\cal A}^{{\rm sl},M_1}_{\rm red}$ defined via
\begin{align}
    {\cal A}^{{\rm sl}, M_1}_{\text{non-rad}} &= \frac{G_F}{\sqrt{2}} V_{ub} C_{\rm sl} \, \langle M_1 \ell^- \bar{\nu}_\ell | \mathcal{O}_{\rm sl} | \bar{B} \rangle \nonumber \\
    &\equiv \frac{G_F}{\sqrt{2}}\,  V_{ub} \, 4E_{M_1} \, [\bar{u}(p_\ell) \frac{\slashed{n}_-}{2} (1-\gamma_5) v_{\nu_\ell}(p_\nu) ] \, {\cal A}^{{\rm sl}, M_1}_{\text{red}} \,.
\end{align}
Here  ${\cal O}_{\rm sl} = \left[\bar{u}\gamma^\mu (1-\gamma_5)b\right]  \left[\bar{l}\gamma_\mu(1-\gamma_5)\nu \right]$ 
is the SM operator for the $b \to u \ell^- \bar{\nu}_\ell$ transition, which in QED also receives short-distance corrections from the electroweak scale and therefore has a non-trivial Wilson coefficient $C_{\rm sl}$~\cite{Sirlin:1981ie}. 

A corresponding QED-generalized factorization formula for heavy-light final states, mediated by the respective operators in Eq.~\eqref{eq:Q12} for $b \to c$ transitions, has been derived in~\cite{Beneke:2021jhp}. 
In particular the decays $\bar{B} \to D^+ M^-$ are dominated by the color-allowed tree topologies and are among the theoretically cleanest non-leptonic decays.
Adopting the traditional counting scheme $z \equiv m_c^2/m_b^2 \sim \mathcal{O}(1)$, the spectator-scattering term in the factorization formula is absent, and one obtains\footnote{In this convention, the physical form factor $\mathcal{F}^{BD}$ needs to be divided by the factor $(C_{\rm sl} Z_\ell)$ to coincide with the definition in~\cite{Beneke:2021jhp}.}
\begin{eqnarray}
\label{eq:non-leptonics-DL}
\left\langle D^{+} M^- | {\cal O}_2 |\bar B\right\rangle &=& 4i E_D E_M \, f_{M}
\mathcal{F}^{BD}(m_M^2) \, \int_0^1 du \, 
T(u,z) \Phi_{M}(u) 
 \, ,
\end{eqnarray}
with an appropriately defined physical form factor $\mathcal{F}^{BD}$ in terms of the semi-leptonic $\bar{B} \to D^+ \ell^- \bar{\nu}_\ell$ amplitude. 
Note that contributions from the operator ${\cal O}_1$ would be of order $\mathcal{O}(\alpha_s \alpha)$, and are neglected.

\subsection{QED generalized light-cone distribution amplitudes}

\label{subsec:QED_LCDAs}

The conventional light-cone distribution amplitudes in QCD are defined by meson-to-vacuum matrix elements of light-ray operators, see the examples given in Eqs.~(\ref{eq:intro:pionLCDA}) and (\ref{eq:intro:BLCDA}) above.
On the other hand, a consistent definition of generalized LCDAs that appear in QED factorization theorems for non-radiative amplitudes requires to specify how the 
real-photon radiation should be taken into account in physical observables.
We emphasize that in the following we consider the case where the experimental resolution energy $\Delta E$ is well below the strong interaction scale $\Lambda$, and the generalized LCDA thus contain virtual QED effects below a soft reference scale $\mu_0$ and above some IR scale $\mu_{\rm IR}>\Delta E$, see also Table~\ref{tab:scales}.
A proper definition then implies a prescription for subtracting the IR divergences appearing in the limit $\mu_{\rm IR} \to 0$, which can be achieved by considering them as IR-finite non-perturbative matching coefficients for an effective theory of ultrasoft radiation with point-like coupling to mesons.
In the following, we summarize the generic properties and novel features of these functions following from the renormalization of the defining non-local operators.

\subsubsection{Light-meson LCDAs}

The QED-generalized leading-twist LCDAs for charged light pseudoscalar mesons has been first introduced in~\cite{Beneke:2020vnb} to include virtual collinear photon exchanges in the QED factorization for non-leptonic $B$ decays. 
For instance, for a charged pion, it can be defined as 
\begin{eqnarray}
\label{eq:lightLCDA}
 \langle \pi^-(p)| R_c^{(Q_\pi)} 
\bar{\chi}^{(d)}(sn_+) \, \slashed{n}_+ \gamma_5 \,\chi^{(u)}(0)| 0\rangle
&=& -i f_\pi \, (n_+\cdot p) \int_0^1 \! du\, e^{i u (n_+\cdot p) s} \, \Phi_{\pi}(u;\mu) \,.
\cr && 
\hspace{1em}
\end{eqnarray}
Here, the decay constant $f_\pi$ is chosen to be the renormalization-scale independent decay constant in QCD \emph{without} QED corrections, and 
the above-mentioned IR subtraction is understood implicitly. 
The structure of the Wilson lines in the definition~\eqref{eq:lightLCDA} is determined from gauge invariance. It is noteworthy that, in contrast to QCD, the QED part of the Wilson lines from~\eqref{eq:QEDquarkblocks} does no longer combine to a finite-length gauge link.
Instead, writing $W^{(Q_d)}(sn_+) W^{(Q_u) \dagger}(0) = [sn_+,0]^{(Q_d)} W^{(Q_\pi)}(0)$, the operator contains a Wilson line with the total charge $Q_\pi = -1$ of the pion extending to $-\infty$, in addition to the finite segment $[sn_+,0]^{(Q_d)}$ with charge $Q_d$.
As a consequence, the collinear operator in~\eqref{eq:lightLCDA} needs to be multiplied with the same kind of
soft-rearrangement factor $R_c^{(Q_\pi)}$ as in~\eqref{eq:Rminus} 
to remove the IR divergences from the anomalous dimension, cf. footnote~\ref{footnote:conventions}.

The renormalization properties of $\Phi_\pi(u;\mu)$ have been studied in detail in~\cite{Beneke:2021pkl}. The evolution kernel 
\begin{align}
\label{eq:lightLCDAanom_dim}
  \Gamma(u,v;\mu) = &- \frac{\alpha_s C_F + \alpha Q_{q_1} Q_{q_2}}{\pi} \, V(u,v) \nonumber  \\ &- \frac{\alpha}{\pi} \, \delta(u-v) \, Q_M\left( Q_M \left( \ln \frac{\mu}{2E} + \frac34 \right) - Q_{q_1} \ln {u} + Q_{q_2} \ln \bar{u} \right)
\end{align}
has been derived for charged ($q_1=d, q_2 = u, Q_M = -1$) and neutral ($Q_{q_1}=Q_{q_2}=Q_q, Q_M = 0$) mesons.
Here $V(u,v)$ is the ERBL kernel from~\eqref{eq:ERBL},
which now receives QED corrections proportional to the product of quark charges. 

The most interesting features show up for charged mesons, from the 
additional local contributions in the second line of~\eqref{eq:lightLCDAanom_dim}. 
Notably, these novel terms have a logarithmic dependence on the large recoil energy of the meson and they feature additional isospin-breaking logarithms of the momentum fractions $u$ and $\bar{u}$ proportional to the respective quark charges. 
The explicit energy dependence can be viewed as a manifestation of the so-called collinear anomaly in SCET~\cite{Becher:2010tm}, which results from the breaking of boost-invariance through the soft rearrangement. 
Being independent of the quark charges, this contribution cancels in ratios with the collinear matrix element $Z_\ell$ of a pointlike lepton, e.g. in~\eqref{eq:non-leptonics-LL} or~\eqref{eq:non-leptonics-DL} when inserting~\eqref{eq:FAred}. We note that, besides this universal energy dependence, the QED generalized LCDA for light mesons retains its universality, and is relevant to various kinds of hard exclusive processes.

The isospin-breaking logarithms $\ln u$ and $\ln \bar{u}$ render the LCDA slightly asymmetric due to the different charges of up- and down-type quarks. They further result in deviations from the standard linear endpoint behaviour for $u\to 0$ and $u \to 1$, dictated by conformal symmetry in QCD. In fact, for asymptotically high scales, the QED generalized LCDA would even diverge at the endpoints. 
However, for the factorization scales appearing in any phenomenological application the inverse moments $\langle u^{-1}\rangle$ and $\langle \bar u^{-1}\rangle $ still exist, 
and the convolutions in the QED factorization theorems remain well-defined.

It is again useful to perform the Gegenbauer expansion
\begin{equation}
\label{eq:Gegenbauer}
    \frac{\Phi_M(u;\mu)}{Z_\ell(\mu)} = 6u \bar{u} \sum_{n=0}^\infty a_n(\mu) \, C_n^{\left( 3/2 \right)}(2u-1) \,,
\end{equation}
where the collinear function $Z_\ell$ has been used as a normalization to absorb the universal energy-dependent logarithms.
Notice that the extra terms in the 1-loop evolution kernel~(\ref{eq:lightLCDAanom_dim}) now induce a mixing between the 
individual Gegenbauer coefficients $a_n$, and due to the isospin-breaking QED effects, also odd coefficients contribute.

Performing a fixed-order expansion in the electromagnetic coupling, 
\begin{equation}
\label{eq:Gegenbauer_FO}
    a_n(\mu) = a_n^{\rm QCD}(\mu) + \frac{\alpha(\mu)}{\pi} \, a_n^{(1)}(\mu) + \mathcal{O}(\alpha^2) \,,
\end{equation}
an analytic solution that resums the leading QCD logarithms for the linear QED terms has been derived in~\cite{Beneke:2021pkl},
\begin{eqnarray} \label{eq:QEDGegenbauer}
   a_n^{(1)}(\mu) &=&  \left(\frac{\alpha_s(\mu)}{\alpha_s(\mu_0)}\right)^{\frac{C_F \gamma_n}{\beta_0}}a_n^{(1)}(\mu_0) -\frac12 Q_{q_1} Q_{q_2} \gamma_n \, a_n^{\rm QCD}(\mu) \, \ln \frac{\mu}{\mu_0} \nonumber \\ 
&&\hspace*{-1.4cm}-  \,Q_M \sum_{m=0}^\infty \frac{2\pi  f_{nm} }{\beta_0^{\rm QCD}+(\gamma_n-\gamma_m)C_F} \left\{ \frac{a_m^{\rm QCD}(\mu)}{\alpha_s(\mu)} - \frac{a_m^{\rm QCD}(\mu_0)}{\alpha_s(\mu_0)}  \left( \frac{\alpha_s(\mu)}{\alpha_s(\mu_0)}  \right)^{\frac{C_F \gamma_n}{\beta_0}} \right\} .
\cr &&
\end{eqnarray}
A particularly interesting feature is the fact that even the
Gegenbauer coefficient $a_0(\mu)$ has a non-trivial RG evolution in QED, and also participates in the mixing. This implies that the QED LCDA is no longer normalized at all scales, $\int_0^1 \! du \, \Phi_M(u;\mu) \neq 1$.

An analytic form of the mixing matrix $f_{nm}$ can be found in (45) of~\cite{Beneke:2021pkl}.
For a negatively charged meson, the first few entries read
\begin{equation} \label{eq:fnmmatrix}
    f = \begin{pmatrix} 
    \frac56 & \frac14 & \frac{3}{10} &  \dots \\[2pt]
    \frac{5}{36} & \frac{31}{30} & \frac{5}{18} & \\[2pt]
    \frac{7}{60} & \frac{7}{36} & \frac{473}{420} & \\
    \vdots & & & \ddots 
    \end{pmatrix} \,.
\end{equation}
Notice that the matrix $f_{nm}$ is not triangular, and therefore higher Gegenbauer moments mix into lower ones and vice versa (in contrast to 2-loop QCD effects). As the entries $f_{n\neq m}$ fall of 
sufficiently fast, and the overall effect of the QED corrections is small anyway, for phenomenological applications it is still legitimate to truncate the Gegenbauer expansion, say at $n=2$ or $n=4$, and to assume that the odd Gegenbauer coefficients vanish at the soft reference scale.
In this way, the numerical impact of structure-dependent QED corrections on the first inverse moments of the pion LCDA turns out to be at the percent level.

\subsubsection{$B$-meson light-cone distribution amplitudes aka soft functions}
\label{sec:QED-B-LCDA}
The generalization of the leading-twist $B$-meson LCDA was first introduced in the study of structure-dependent QED corrections to the decay $\bar{B}_q \to \mu^+ \mu^-$ in~\cite{Beneke:2019slt}.
As a consequence of the non-decoupling of soft photons, this generalized object possesses knowledge about the charges and directions of the energetic charged particles in the final state. 
In contrast to the generalized collinear LCDAs above, the $B$-meson LCDA thus turns into a process-specific soft hadronic function 
which also includes soft QED rescattering phases.

A definition that holds for all possible charge combinations relevant in two-body back-to-back decays has been introduced in~\cite{Beneke:2020vnb},
\begin{align}
\label{eq:QED_BLCDA}
&\frac{\langle 0| \bar{q}_s(tn_-) Y^{(Q_q)}(tn_-,0)  
\,\slashed{n}_-\gamma_5 \, h_v(0) Y_{+}^{(Q_{1}) \dagger}(0) Y_{-}^{(Q_{2}) \dagger}(0) |\bar{B}_v\rangle}{R_c^{(Q_{1})}R_{\bar{c}}^{(Q_{2})}} \nonumber  \\ 
 = \, &i F_{\rm stat}(\mu) \int_{-\infty}^\infty d\omega\; e^{-i \omega t} \,\Phi_{B, \otimes}(\omega;\mu) \, ,
\end{align}
where the $\otimes$ symbol again labels the charge combinations $(Q_{1}, Q_{2})$ of the final state. Similar to the definition~\eqref{eq:lightLCDA} of the pion LCDA, $F_{\rm stat}(\mu)$ has been \emph{chosen} to coincide with the static $B$-meson decay constant in the absence of QED, such that all additional QED effects are captured by the functions $\Phi_{B, \otimes}(\omega;\mu)$. %

Notice that -- in contrast to the conventional LCDAs in QCD -- the non-local operator defining the soft function $\Phi_{B, \otimes}(\omega;\mu)$ 
involves fields separated along two \emph{different} light-cone directions, $n_+$ and $n_-$, as a consequence of the non-local Wilson line $Y_{-}^{(Q_2) \dagger}$.
Moreover, as a novel feature, the soft function has support on the \emph{entire} real axis, $\omega\in (-\infty,\infty)$.
While this may be surprising at first glance, it can be traced back to the presence of the Wilson line $Y_{-}^{(Q_{2}) \dagger}$ 
which arises from a charged particle moving in the anti-collinear direction with large light-cone momentum component $(n_-\cdot q) \sim \mathcal{O}(m_b)$.
Soft-photon couplings to the outgoing Wilson line can thus reduce the light-cone momentum component $\omega = (n_- \cdot k_s)$ of the light quark in
the $B$-meson by an infinite amount in the heavy-quark limit.

The extended support for $\omega <0$
greatly complicates the renormalization of the soft function~\cite{Beneke:2022msp}.
Besides the standard plus-distributions familiar from the Lange-Neubert evolution kernel in~\eqref{eq:LNkernel}, 
one has to define two additional modified plus-distributions,
\begin{align} 
 \int_{-\infty}^\infty d\omega' \big[ \dots \big]_{\oplus / \ominus} \,\phi(\omega') &= 
 \int_{-\infty}^\infty d\omega' \big[ \dots \big] \big(\phi(\omega')-\theta(\pm \, \omega') \, \phi(\omega) \big) \,.
\end{align}
Defining the auxiliary distributions
\begin{align}
\begin{aligned}
\label{eq:FandG1}
    {F}^>(\omega,\omega') &= \omega \left[ \frac{\theta(\omega'-\omega)}{\omega'(\omega'-\omega)} \right]_+ + \left[ \frac{\theta(\omega-\omega')}{\omega-\omega'}\right]_{\oplus} \,, \\   
    {F}^<(\omega,\omega') &=  \omega \left[ \frac{\theta(\omega-\omega')}{\omega'(\omega-\omega')} \right]_++  \left[\frac{\theta(\omega'-\omega)}{\omega'-\omega} \right]_{\ominus} \,,
    \end{aligned}
\end{align}
and
\begin{align}
\begin{aligned}
\label{eq:FandG2}
    {G}^>(\omega,\omega') &= \left(\omega+\omega'\right) \left[ \frac{\theta(\omega'-\omega)}{\omega'(\omega'-\omega)} \right]_+ -i\pi \,\delta(\omega-\omega') \,, \\
    {G}^<(\omega,\omega') &= \left(\omega+\omega'\right) \left[ \frac{\theta(\omega-\omega')}{\omega'(\omega-\omega')} \right]_+ +i\pi \,\delta(\omega-\omega') \,,
    \end{aligned}
\end{align}
the evolution kernel for $\Phi_{B,\otimes}$ can be expressed in a compact form in terms of the linear combinations 
\begin{align}
\label{eq:Hpm}
    H_\pm(\omega, \omega') \equiv \theta(\pm\omega)\, F^{> (<)} (\omega,\omega')+ \theta(\mp\omega)\, G^{< (>)}(\omega,\omega')
\end{align}
as
\begin{align} 
\label{eq:QED_BLCDA_evolution}
    \Gamma_\otimes(\omega,\omega')  &= \frac{\alpha_s C_F}{\pi} \bigg[ \left(\ln \frac{\mu}{\omega-i0} - \frac12 \right)\delta(\omega-\omega') -H_+(\omega,\omega') \bigg] \nonumber \\ 
    &+ \frac{\alpha}{\pi} \bigg[  \bigg(  (Q_s^2 +2Q_s Q_{1}) \ln \frac{\mu}{\omega-i0}  - \frac34 Q_s^2 -\frac12 Q_d^2  \\&+ i\pi (Q_s + Q_{1}) Q_{2} \bigg) \delta(\omega-\omega') - Q_s Q_d H_+(\omega,\omega') +  Q_s Q_{2} H_-(\omega,\omega') \bigg] \,, \nonumber
\end{align}
which holds for general two-body back-to-back decays. 
Here $Q_s$ is the electric charge of the $B$-meson spectator-quark field in~\eqref{eq:QED_BLCDA}. The superscripts $>(<)$ in~Eqs.~(\ref{eq:FandG1},\ref{eq:FandG2}) indicate whether the distributions act on a test functions with support $\omega > 0$ ($\omega < 0$), and it is noteworthy that these two branches mix under renormalization. Negative support is only generated for $Q_2 \neq 0$, and in that case the logarithm of $\omega - i0$ gives rise to a complex phase.
On the other hand, for the case $Q_2 = 0$, the distribution $H_+$ reduces to the expression in the Lange-Neubert kernel,
\begin{equation}
    H_+(\omega,\omega') \to 
    \omega \left( \left[ \frac{ ~\theta(\omega'-\omega)}{\omega'(\omega'-\omega)}\right]_+ + \left[\frac{\theta(\omega-\omega')}{\omega(\omega-\omega')}\right]_+ \right) \,.
\end{equation}

The non-vanishing support for $\omega < 0$ and the associated RG evolution also require to retain the $i0$ prescription from the hard-collinear propagators in the 
definition of the inverse moments,
\begin{align} 
\label{eq:BLCDA_invmoment}
    \frac{1}{\lambda_{B}(\mu)} &= \int_{-\infty}^\infty \frac{d\omega}{\omega-i0} \,\Phi_{B,\otimes}(\omega;\mu) \,,
\end{align}
since for $Q_2\neq 0$ the soft function does not vanish at $\omega = 0$.
In QED factorization theorems, this yields an additional source of soft imaginary parts from the integration over the point $\omega = 0$.
In contrast, in QCD factorization 
theorems for exclusive $B$ decays, 
complex phases only arise from hard and hard-collinear loop corrections.
Despite the complexity of the evolution kernel~\eqref{eq:QED_BLCDA_evolution}, the first inverse moment $\lambda_B^{-1}(\mu)$ at the scale $\mu$ can again 
be expressed in terms of the soft function at a reference scale $\mu_0$ in an analogous way as in Eq.~(\ref{eq:invmoment_evol_QCD}),
\begin{align}
 \lambda_B^{-1}(\mu) &= e^{{\sf V}+2\gamma_E {\sf a}} \, \frac{\Gamma(1+{\sf a})}{\Gamma(1-{\sf a})} \, \mathcal{F}(0;\mu,\mu_0) \, 
 \int_{-\infty}^\infty \frac{d\omega}{\omega-i0} \left( \frac{\mu_0}{\omega-i0} \right)^{\!{\sf a}} \Phi_{B}(\omega;\mu_0) \,,
\end{align}
with the QED-generalized evolution functions \begin{align}
 {\sf  a} = {\sf a}(\mu,\mu_0) &= a(\mu,\mu_0) - P_Q \int_{\alpha(\mu_0)}^{\alpha(\mu)} \frac{d\alpha}{\beta_{\rm em}(\alpha)} \, \frac{\alpha}{\pi} \,, \qquad P_Q = Q_s  (2 Q_{1} + Q_s) 
 \end{align}
 and 
 \begin{align}
  {\sf V} = {\sf V}(\mu,\mu_0)
  &= V(\mu,\mu_0) -P_Q \int_{\alpha(\mu_0)}^{\alpha(\mu)} \frac{d\alpha}{\beta_{\rm em}(\alpha)} \, \frac{\alpha}{\pi} \int_{\alpha(\mu_0)}^{\alpha} \frac{d\alpha'}{\beta_{\rm em}(\alpha')}\nonumber \\[0.2em] 
  &\qquad {} -\int_{\alpha(\mu_0)}^{\alpha(\mu)} \frac{d\alpha}{\beta_{\rm em}(\alpha)} \, \frac{\alpha}{\pi} \bigg[ - \frac34 Q_s^2 - \frac12  Q_d^2 + \,i\pi (Q_s +Q_{1} ) Q_{2} \bigg]\,, 
  \end{align}
  together with 
  \begin{align}
  \mathcal{F}(\eta;\mu,\mu_0) &= \exp \bigg\{ \int_{\mu_0}^{\mu} \frac{d\mu'}{\mu'}  \frac{\alpha(\mu') \, Q_s Q_{1}}{\pi} (H_{\eta+{\sf a}(\mu,\mu')} + H_{-\eta-{\sf a}(\mu,\mu')}) \bigg\} \,. 
\end{align}
Here $H_x = \gamma_E+ \psi(x+1)$ is the harmonic number function.
Numerically, the size of QED corrections to these inverse moments in two-body back-to-back decays has been estimated to be at most $1.2$\%.

The detailed study of the renormalization-group evolution of the soft function itself is quite involved, and would go beyond the scope of this review. A comprehensive analysis can be found in the original article~\cite{Beneke:2022msp}.
It is worth noting that complex-valued soft functions, defined through operators that contain fields that are separated along  different light-cone directions, 
will become relevant in various processes, in particular for the analysis of power corrections in the $1/m_b$ expansion.
In a recent example, such an HQET matrix element has been analyzed in the context of charm-loop contributions to the radiative decay $\bar{B}_{d,s} \to \gamma\gamma$~\cite{Qin:2022rlk}.
Interestingly, it has been shown that the support of the appearing three-particle soft function also extends to the negative axis, and the corresponding evolution kernel features the very same distributions $H_\pm$ as in~\eqref{eq:Hpm}.


\section{QED-induced  rapidity logarithms}

\label{sec:rapidity}
QED effects in exclusive $B$-meson decays are also interesting because they provide a perturbative framework to study the systematics of rapidity logarithms that are related to
endpoint-divergent convolution integrals in the factorization approach.
The systematic understanding of such endpoint logarithms in the context of 
EFTs is currently a very active field of research, and substantial progress has recently been made in different contexts, see~\cite{Cornella:2022ubo,Boer:2018mgl,Feldmann:2022ixt,Hurth:2023paz,Bell:2023hzm} in the field of $B$-physics, and e.g.~\cite{Liu:2019oav,Beneke:2020ibj,Bell:2022ott} for other applications.

\subsection{$\bar B_s \to \mu^+\mu^-$}

\label{sec:O7}

An example where non-local QED corrections lead to endpoint-divergent convolution integrals in the factorization approach, 
namely the power-enhanced contribution of the electromagnetic penguin operator ${\cal O}_7$
in the $B_s \to \mu^+\mu^-$ decay amplitude, has already been mentioned above.
Here, the appearance of rapidity logarithms is related to the internal muon propagator which allows for a perturbative QED analysis, entangled with the strong dynamics of the spectator quark in the $B$-meson. 
In the following, we briefly sketch the main analytical results and conceptual insights, following the original analyses in \cite{Beneke:2017vpq,Beneke:2019slt} and 
a dedicated work on rapidity divergences in $\bar B_s \to \mu^+\mu^-$ in \cite{Feldmann:2022ixt}.

One identifies the form factor multiplying the Wilson coefficient $C_7^{\rm eff}$ in (\ref{eq:dimuon:amp}) as 
\begin{eqnarray} \label{eq:C7FLO}
{\cal F}^{\rm LO} (E,m) &=& \int_0^\infty \frac{d\omega}{\omega} \, \phi_B^+(\omega) 
    \left[ \frac12 \, \ln^2 \frac{m^2}{2E\omega} +  \ln \frac{m^2}{2E\omega} + \frac{\pi^2}{3} \right] \,.
\end{eqnarray}
Decomposing the box diagram into different momentum regions, one finds the following structure of the \emph{bare} factorization theorem
\cite{Feldmann:2022ixt},
\begin{eqnarray}
{\cal F}(E,m) &=& \int_0^\infty \frac{d\omega}{\omega} \, \phi_B^+(\omega) 
\Bigg\{ \int_0^1 \frac{du}{u} \, H_1(u) \, \bar J_1(u;\omega) \,
\cr 
&& \qquad {} + \bar J_2(1,\omega) \, \int_0^1 \frac{du}{u} \, H_1(u) \, \bar C(u;\omega) 
\cr 
&&
\qquad {} + H_1(0) \, \int_0^\infty \frac{du}{u} \, \int_0^\infty \frac{d\rho}{\rho} \, S(u,\rho;\omega) \, \bar J_2(1+\rho,\omega)  \Bigg\}_{\rm bare} ,
\label{eq:bare}
\end{eqnarray}
where 
\begin{eqnarray}
H_1(u) & = & 1 + {\cal O}(\alpha_s) \, , \qquad 
\bar J_2(z) =  \frac{1}{z} + {\cal O}(\alpha_s) 
\end{eqnarray}
and $u,\rho$ denote the light-cone momentum fractions of the internal muon.
The 1-loop expressions arising from the integration over hard-collinear, collinear and soft momentum regions take the following form,
\begin{eqnarray}
\bar J_1(u;\omega) & = & -  \Gamma(\epsilon)  \left( \frac{\mu^2 e^{\gamma_E}}{2E\omega u(1-u)} \right)^\epsilon (1-u) + {\cal O}(\alpha_s) \, , \\
\bar C^{(1)}(u;\omega)  
    &=& 
	 \Gamma(\epsilon) \,
\left(\frac{\mu^2 e^{\gamma_E}}{m^2}\right)^{\epsilon}
		\left( \frac{\nu^2}{2E\omega} \right)^\delta
	\left( 1-u \right)^{1-2\epsilon} \, u^{-\delta}  + {\cal O}(\alpha_s)\,,
 \\
 S^{(1)}(u,\rho;\omega) &=& \theta(u\rho -\lambda^2) \left( \frac{\mu^2  e^{\gamma_E}}{2E\omega} \right)^\epsilon \left( \frac{\nu^2}{2uE\omega} \right)^\delta   \frac{(u\rho -\lambda^2)^{-\epsilon}}{ \Gamma(1-\epsilon)}  + {\cal O}(\alpha_s)\, \,.
\end{eqnarray}
Here $\epsilon=D/2-2$ refers to dimensional regularization, whereas the soft and collinear momentum regions require to introduce an additional analytic regulator $\delta$, because otherwise the convolution integrals in the bare factorization theorem would be ill-defined.\footnote{For the explicit definition of the regulator, see Ref.~\cite{Feldmann:2022ixt}.}
The key feature that allows one to get rid of the analytic regulator
in the bare factorization theorem, is the presence of so-called refactorization conditions 
\cite{Boer:2018mgl,Liu:2019oav}.
In the present case, the endpoint behaviour of the collinear function
can be written as 
\begin{eqnarray}
&& \left[\left[\bar C(u;\omega)\right]\right] \equiv
\bar C(u;\omega)\big|_{u\to 0}
= \int_0^\infty \frac{d\rho}{\rho} \, S(u,\rho;\omega) + {\cal O}(\alpha) \,,
\label{refac}
\end{eqnarray}
which is explicitly fulfilled by the 1-loop expressions above.
Furthermore, one uses  
that for a suitably chosen analytic regulator, the following double integral is scaleless,
\begin{eqnarray}
\int_0^\infty \frac{du}{u} \, \int_0^\infty \frac{d\rho}{\rho} \, S(u,\rho;\omega) &=& 0 \,.
\end{eqnarray}
The bare factorization theorem can then be rewritten as
\begin{eqnarray}
{\cal F}(E,m) &=& \int_0^\infty \frac{d\omega}{\omega} \, \phi_B^+(\omega) 
\Bigg\{ \int_0^\infty \frac{du}{u} \Big[ H_1(u) \, \bar J_1(u;\omega) \, \theta(1-u) 
\cr && \qquad \qquad \qquad \qquad   {}
- H_1(0) \, \bar J_2(1,\omega) \, \theta(u-1) \, \int_1^\infty\frac{d\rho}{\rho} \, S(u\rho;\omega) \Big]_{\lambda^2 \to 0} 
\cr 
&& \qquad {} + \bar J_2(1,\omega) \, \int_0^1 \frac{du}{u} \Big[ H_1(u) \,  \bar C(u) - H_1(0) \left[\left[ \bar C(u)\right]\right] \Big]
\nonumber\\[0.5em]
&& \qquad {} +H_1(0) \, \int_0^\infty \frac{du}{u} \, \int_0^\infty \frac{d\rho}{\rho} \left[ \bar J_2(1+\rho,\omega) -  
\theta(1-\rho) \, \bar J_2(1,\omega) \right] S(u \rho;\omega)
\nonumber\\[0.3em]
&&  \qquad {} + \bar J_2(1,\omega) \, H_1(0) \, \int_0^1 \frac{du}{u}\, \int_0^1 \frac{d\rho}{\rho} \,  S(u \rho;\omega) \Big|_{\rm \lambda^2\to 0} \Bigg\} \,,
\label{fact}
\end{eqnarray}
where now all endpoint-divergent contributions are subtracted on the level of the integrands.
The limit $\delta \to 0$ can thus be performed \emph{before} the integrations, and as a consequence the collinear function does not depend on $\omega$ anymore, and the soft function only depends on the product $u\rho$ (and, via the dimensional regulator, on $\omega$).
Notice that the emerging cut-offs in the first and fourth term lead to additional power corrections in $\lambda^2$ which should be dropped, as indicated by the limit $\lambda^2 \to 0$.
In this form of the factorization theorem, the double-logarithmic term in Eq.~(\ref{eq:C7FLO}) solely arises from the last term, and it is a straight forward exercise to implement the leading logarithmic QCD corrections from the scale evolution of the individual terms, see Ref.~\cite{Feldmann:2022ixt}.
On the other hand, \emph{additional} QED corrections
would also modify the refactorization condition~(\ref{refac}), and the QED corrections to the $B_s$-meson LCDA, discussed in~\ref{sec:QED-B-LCDA}, have to be taken into account as well.

\subsection{$B^- \to \mu^- \bar \nu_\mu $}

Recently, it has also been shown how to treat structure-dependent QED effects in the charged-current leptonic decay of $B$-mesons, when virtual photon exchanges 
between the light quark in the $B$-meson and the muon are taken into account \cite{Cornella:2022ubo}.
Due to the $(V-A)$ structure of the effective operator, no power enhancement of non-local QED corrections appear in this case.
Similarly as in the previous subsection, the bare factorization theorem for the separation of the various momentum modes in SCET suffers from 
endpoint-divergent convolution integrals. 
This issue can again be resolved by performing 
suitable subtractions along the lines developed in \cite{Liu:2019oav}.
An important observation made in \cite{Cornella:2022ubo} is that the rearrangement to remove the spurious IR-dependence of the anomalous dimensions is then obsolete. 
In particular, this requires a modification of the $B$-meson decay constant in the EFT,
\begin{align}
    \langle 0|{\cal O}_A^{(\Lambda_F)}|B^-(v)\rangle
    &= i m_B \, F(\mu,\Lambda_F) \, \langle 0|Y_v^{(B)} Y_+^{(\ell)\dagger}|0\rangle \,,
\end{align}
which involves a vacuum matrix element of a light-like and a time-like soft Wilson line, and now depends on a subtraction scale $\Lambda_F$ via the non-local operator
\begin{align}
  {\cal O}_A^{(\Lambda_F)} &= \bar q_s(0) \, \slashed n_-\gamma_5 \, h_v(0) \, \theta(i n_+ D_s - \Lambda_F) \, Y_+^{(\ell)\dagger} \,.
\end{align} 
In this definition 
the subtraction parameter $\Lambda_F$ 
in the gauge-covariant form of the $\theta$-function constrains 
the light-cone momentum of the photon momenta in the QED Wilson line. 
The renormalization-scale dependence of the generalized decay constant at 1-loop now also receives QED contributions via
\begin{align}
  \frac{d\ln F}{d\ln\mu } &=- \frac{3C_F\alpha_s}{4\pi} + \frac{3\alpha}{4\pi}\left( Q_\ell^2-Q_b^2+\frac23 \, Q_\ell Q_u \, \ln \frac{\Lambda_F^2}{\mu^2} \right) \,,
\end{align}
whereas the dependence on the subtraction scale $\Lambda_F$ involves a logarithmic moment of the subleading $B$-meson LCDA,
\begin{align}
    \frac{d\ln F}{d\ln\Lambda_F} &= Q_\ell Q_u \, \frac{\alpha}{2\pi} \left[ \int_0^\infty d\omega \, \phi_B^-(\omega) \, \ln \frac{\omega \Lambda_F}{\mu^2} -1 +\ldots\right] ,
\end{align}
where the dots stand for additional terms involving a combination of 3-particle LCDAs of the $B$-meson.\footnote{For additional subtleties arising at $\mathcal{O}(\alpha \alpha_s)$, see the discussion in~\cite{Cornella:2022ubo}.}
Choosing $\Lambda_F=m_b$ for simplicity, the amplitude for $B^-\to \mu^-\bar\nu_\mu$ in the presence of virtual QED corrections can then 
be written as
\begin{align}
    {\cal A}^{\rm virtual}_{B\to \mu \bar\nu_\mu} &= 
    i \sqrt2 \, G_F K_{\rm EW}(\mu) V_{ub} \frac{m}{m_b} \, \bar u(p_\mu) \, P_L \, v(p_\nu) 
    \cdot m_B F(\mu,m_b) \left[{\cal M}_{2p} + {\cal M}_{3p} \right] ,
\end{align}
where the contribution from the 2-particle Fock state in the $B$-meson is calculated as
\begin{align}
    {\cal M}_{2p}(\mu) &= 1 + \frac{\alpha_s C_F}{4\pi} \left[ \frac32 \, \ln \frac{m_b^2}{\mu^2}-2 \right]
    + \frac{\alpha}{4\pi} \Bigg\{ Q_b^2 \left[\frac32 \, \ln \frac{m_b^2}{\mu^2} -2 \right] \cr 
    & \quad {}
    - Q_\ell Q_b \left[\frac12 \, \ln^2 \frac{m_b^2}{\mu^2} + 2 \, \ln \frac{m_b^2}{\mu^2} - 3 \, \ln \frac{m^2}{\mu^2}+1+\frac{5\pi^2}{12} \right]
    \cr 
    & \quad {} + 2 Q_\ell Q_u \, \int_0^\infty d\omega \, \phi_B^-(\omega) \, \ln \frac{m_b \omega}{\mu^2} 
    \cr & \quad {}
    + Q_\ell^2 \left[ \frac{1}{\epsilon_{\rm IR}} \left( \ln \frac{m_B^2}{m^2} - 2 \right) 
    + \frac12 \, \ln^2 \frac{m^2}{\mu^2} - \frac12 \, \ln \frac{m^2}{\mu^2} + 2 + \frac{5\pi^2}{12} \right] \Bigg\}\,,
\end{align}
and the 3-particle contribution ${\cal M}_{3p}$ can be found in \cite{Cornella:2022ubo}. 
The factor $K_{\rm EW}(\mu)$ takes into account short-distance electroweak corrections 
to the charged-current operator, with the 1-loop scale dependence given by 
\begin{align}
    \frac{dK_{\rm EW}}{d\ln\mu} &= Q_\ell Q_u \, \frac{3\alpha}{2\pi} \, K_{\rm EW}(\mu)\,.
\end{align}


\section{Summary}

In anticipation of the high-precision measurements from LHCb
and BELLE II, the investigation of electromagnetic effects in $B$-meson decays has become an active research field in recent years
(besides the articles which have already been mentioned in the main text, more examples can be found in Refs.~\cite{Bigi:2023cbv,Choudhury:2023uhw,Isidori:2022bzw,Nabeebaccus:2022jhu,
Gorbahn:2022rgl,Papucci:2021ztr,Mishra:2020orb,Cali:2019nwp,deBoer:2018ipi,Huber:2015sra,Carrasco:2015xwa}).
In this review, we have summarized recent developments in the calculation of \emph{structure-dependent} QED corrections to exclusive $B$-meson decays with effective field theory methods.
More precisely, we have focused on the factorization of virtual QED corrections to the non-radiative decay amplitudes between the hadronic scale $\Lambda$ and the $B$-meson mass $m_B$.
A rigorous theoretical treatment of these effects is very intricate and subtle, as the decoupling of soft photons from charged external mesons and leptons is spoiled.
As a consequence, certain infrared divergent terms in the hadronic matrix elements require a special and careful rearrangement, and overall the factorization of QED effects turns out to be more complicated than in QCD. 

Despite the smallness of the electromagnetic coupling, structure-dependent QED effects are interesting for various reasons. 
Phenomenologically, in certain observables their effects can be relevant, and uncertainties from novel hadronic matrix elements must be added to the error budget of a given process.
The numerical impact is particularly prominent for the $\bar{B}_s \to \mu^+ \mu^-$ decay due to the power-enhancement of non-local QED corrections. 
On the other hand, QED corrections also provide a playground to study open field theoretical problems related to factorization at subleading orders in the $1/m_B$ expansion.  
Also the non-perturbative matching of the hadronic matrix elements in SCET onto an effective theory of very low-energetic photons with point-like couplings to the (boosted) hadrons has not been worked out in the literature so far.

The systematic treatment of QED corrections along these lines is a relative new field, and QED factorization theorems for many processes are still unknown. 
A generalization of these methods from the muonic and non-leptonic two-body decays discussed here, e.g. to kinematic distributions in semi-leptonic transitions, or to different lepton species, would certainly be of phenomenological and theoretical interest.
Lastly, we stress the importance of a consistent treatment of QED effects between theoretical predictions and experimental analyses, in particular regarding very low-energetic photons.
We hope that this summary provides a valuable resource for future analyses of QED corrections in $B$ decays.

\subsection*{Acknowledgments}

The research of T.F.\ is supported by the Deutsche Forschungsgemeinschaft (DFG, German Research
Foundation) under grant 396021762 -- TRR 257.
The research of P.B.\ has been supported by funds from the Cluster of Excellence Precision Physics, Fundamental Interactions, and Structure of Matter (PRISMA$^+$ EXC 2118/1) funded by the German Research Foundation (DFG) within the German Excellence Strategy (Project ID 390831469), and from the European Research Council (ERC) under the European Union’s Horizon 2022 Research and Innovation Programme (Grant agreement No.101097780, EFT4jets).

\subsection*{Data Availability Statement}

No data associated in the manuscript.


\bibliographystyle{JHEP}
\bibliography{bib}


\end{document}